 \documentclass[final,5p,times,twocolumn,lefttitle]{elsarticle}

\usepackage{amssymb,amsmath,amsfonts}
\usepackage{natbib}
\usepackage{graphicx}
\usepackage{dcolumn}
\usepackage{tensor}
\usepackage{bm}
\usepackage{float}
\usepackage{color}
\usepackage[dvipsnames]{xcolor}
\usepackage{listings}
\usepackage{fancyvrb}
\usepackage{slashed}
\usepackage{threeparttable}
\usepackage{enumitem}
\usepackage{listings}
\usepackage{mathtools}
\usepackage{slashed}
\usepackage{verbatim}
\usepackage{hyperref}
\hypersetup{colorlinks, citecolor=blue, linkcolor=blue, urlcolor=blue,pdfauthor=author}
\usepackage{braket}

\def\ee{\end{eqnarray}}

\newcommand{\be}{\begin{eqnarray}}
\newcommand{\en}{\end{eqnarray}}
\newcommand{\bea}[1]{\left(\begin{array}{#1}}
\newcommand{\ena}{\end{array}\right)}
\newcommand{\ba}{\begin{eqnarray}}
\newcommand{\ea}{\end{eqnarray}}

\journal{Physics Letters B}

\biboptions{square,comma,compress}

\begin{document}

\begin{frontmatter}



\title{The Gallium Solar Neutrino Capture Cross Section Revisited}

\author[first,second]{W. C. Haxton}
\ead{haxton@berkeley.edu}
\cortext[cor1]{Corresponding author}
\affiliation[first]{organization={Department of Physics, University of California},
            city={Berkeley},
            state={California},
            postcode={94720}, 
            country={USA}}
\affiliation[second]{organization={Lawrence Berkeley National Laboratory},
            city={Berkeley}, 
            state={California},
            postcode={94720},
            country={USA}}

\author[first,third]{Evan Rule\corref{cor1}}
\ead{erule@berkeley.edu}
\affiliation[third]{organization={Los Alamos National Laboratory},
            city={Los Alamos},
            state={New Mexico},
            postcode={87545}, 
            country={USA}}

\begin{abstract}
Solar neutrino flux constraints from the legacy GALLEX/GNO and SAGE experiments continue to influence contemporary global analyses of neutrino properties.  The constraints depend on the neutrino absorption cross sections for various solar sources. Following recent work updating the $^{51}$Cr and $^{37}$Ar neutrino source cross sections, we reevaluate the $^{71}$Ga solar neutrino cross sections, focusing on contributions from transitions to $^{71}$Ge excited states, but also revising the ground-state transition to take into account new $^{71}$Ge electron-capture lifetime measurements and various theory corrections.  The excited-state contributions have been traditionally taken from forward-angle $(p,n)$ cross sections.  Here we correct this procedure for the $\approx 10\%$--$20\%$ tensor operator contribution that alters the relationship between Gamow-Teller and $(p,n)$ transition strengths. Using state-of-the-art nuclear shell-model calculations to evaluate this correction, we find that it lowers the $^8$B and hep neutrino cross sections. However, the addition of other corrections, including contributions from near-threshold continuum states that radiatively decay, leads to an overall increase in the $^8$B and hep cross sections of $\approx 10\%$ relative to the values recommended by Bahcall.   Uncertainties are propagated using Monte Carlo simulations.
\end{abstract}

\begin{keyword}
Solar neutrinos \sep Neutrino capture \sep Gallium \sep Charge-exchange reactions\\
\textit{Report numbers}: LA-UR-24-32851, N3AS-24-043



\end{keyword}

\end{frontmatter}

\section{Introduction}
\label{introduction}
The legacy radiochemical gallium experiments,  GALLEX/GNO  \cite{ANSELMANN1992376,GALLEX:1998kcz,Kaether:2010ag,GNO:2005bds}  and SAGE \cite{SAGE:2009eeu,Gavrin:2019sok}, played an important role in resolving the solar neutrino problem.  The combined results from the $^{37}$Cl \cite{PhysRevLett.20.1205}, Kamiokande \cite{Kamiokande-II:1989hkh}, and the gallium experiments produced a pattern of solar neutrino fluxes that was incompatible with the standard solar model (SSM), even if model parameters were adjusted to change the Sun's core temperature \cite{PhysRevD.49.3622,PhysRevD.50.4749,1995ARA&A..33..459H}.  The apparent need for new physics helped to justify a new and more capable generation of experiments that established that neutrinos are massive and undergo flavor mixing. The GALLEX/GNO and SAGE results continue to have impact today, influencing global analyses of neutrino parameters \cite{Esteban:2020cvm,deSalas:2020pgw,PhysRevD.104.083031} and providing (along with Borexino \cite{zavatarelli20161753} results) one of the few direct constraints on the low-energy pp neutrinos.  

The solar electron-neutrino rates extracted from these experiments depend on cross sections calculated by Bahcall \cite{Bahcall:1997eg} more than twenty-five years ago.  Here we update these cross sections, incorporating corrections similar to those introduced in \cite{PhysRevC.108.035502}, where $^{71}$Ga cross sections for the low-energy neutrino sources $^{51}$Cr and $^{37}$Ar were reevaluated.

The neutrino-source work was undertaken because of the ``gallium anomaly" \cite{Elliott:2023cvh}, the discrepancy between the predicted and observed neutrino rates obtained in four neutrino-source calibrations of the 
GALLEX/GNO and SAGE experiments, which now stands at $\approx 4 \sigma$, due to the recent high-intensity BEST experiment \cite{PhysRevLett.128.232501,PhysRevC.105.065502}.  Given the increased significance of the discrepancy, a reevaluation \cite{PhysRevC.108.035502} of the $^{71}$Ga capture cross sections for $^{51}$Cr and $^{37}$Ar neutrino sources was performed, to ensure that associated theory uncertainties had been properly addressed.   The new work employed the most recent Q value and Particle Data Group couplings, and included radiative corrections and the effects of weak magnetism --- the leading contribution from momentum-dependent weak transition operators.  Excited-state contributions to the cross section have traditionally been estimated from forward-angle $(p,n)$ cross sections \cite{Bahcall:1997eg}.  An important correction affecting the proportionality of neutrino capture and $(p,n)$ scattering comes from a sub-leading tensor operator that contributes only to the latter process.  The size of this correction was determined empirically in \cite{PhysRevC.108.035502} and applied to the two excited states of $^{71}$Ge that contribute to the neutrino-source capture cross sections.  This new work yielded both an updated cross section and a Monte-Carlo uncertainty envelope that confirmed nuclear physics is not responsible for the Ga neutrino-source anomaly.

Here we provide a similar update of $^{71}$Ga solar neutrino cross sections, providing best values and uncertainties. The changes we find are small but affect the various solar neutrino sources differently. First, the solar pp neutrinos, which dominate the $^{71}$Ga rate, are captured almost exclusively through the transition to the  $^{71}$Ge ground state.  As this transition is tightly constrained by the known electron-capture (EC) lifetime of $^{71}$Ge, our update reflects small changes from the inclusion of radiative and weak-magnetism corrections and the use of a more accurate Q value.  The next most important source, the line neutrinos produced in the sun by EC on $^7$Be, are absorbed through the same transitions studied in \cite{PhysRevC.108.035502} (see Fig. \ref{fig:levels}).  There it was found that the corrections to the $^{51}$Cr and $^{37}$Ar cross sections associated with the ground- and excited-state transitions tended to cancel, leading to small overall changes of $\approx$ 2\%.    Cross sections for the third most important source,  the high-energy $^8$B neutrinos, were taken in \cite{Bahcall:1997eg} from uncorrected $(p,n)$ cross sections. Given the impressive improvement in the correspondence between weak and $(p,n)$ transition strengths achieved in \cite{PhysRevC.108.035502} when a tensor amplitude correction was applied (see Fig. 2 in that reference), we include the same correction for the $^{71}$Ga $^8$B cross section. 

\section{Neutrino sources and spectra}
\label{sec:spectra}
The solar neutrino sources we address are listed in Table \ref{tab:fluxes}.  
The EC pep and $^7$Be sources produce discrete lines, while the $\beta$-decay pp, $^8$B, hep, $^{13}$N,
$^{15}$O, and $^{17}$F sources produce neutrinos in a continuous spectrum.   
The line energies and the $\beta$-decay end-point energies (neglecting in the Table solar thermal effects)
are denoted by $E_\nu^\mathrm{max}$.  Also shown are the (unoscillated) fluxes taken from the SSM calculations of Herrera and Serenelli \cite{B23Fluxes}, using the
high-metallicity (high-Z) GS98 composition of \cite{Grevesse:1998bj} and the low-Z (AAG21) of \cite{Asplund:2021}. The tension between these results has been often referenced as the solar metallicity problem.  The third set of fluxes, denoted MB22p, incorporates changes in composition and opacities introduced by Magg et al. \cite{Magg:2022rxb}.   

\begin{table*}
\centering
\caption{Solar neutrino $\beta$-decay and EC sources and their maximum or line energies, respectively, in the absence
of solar thermal effects (see text).
The fluxes are taken from the SSM calculations of \cite{B23Fluxes} using the compositions GS98 \cite{Grevesse:1998bj} (high-Z), AAG21 \cite{Asplund:2021} (low-Z), and MB22p \cite{Magg:2022rxb}. Associated uncertainties are indicated.}
\label{tab:fluxes}
\begin{tabular}{lcccccc}
\hline \hline
 Source & Type & $\begin{array}{c} E_\nu \\ \mathrm{(MeV)} \end{array}$ & GS98  & AAG21  & MB22p & $\begin{array}{c} \mathrm{units} \\ \mathrm{(cm}^{-2}\mathrm{s}^{-1}\mathrm{)} \end{array}$  \\
\hline
 ~& & & & &  ~\\[-.2cm]
pp & $\beta^+$ & $\le$0.420 & 5.96\,(0.6\%) & 6.00\,(0.6\%) & 5.95\,(0.6\%) &  10$^{10}$ \\[0.1cm]
pep & EC & 1.442 &  1.43\,(1.1\%) & 1.45\,(1.1\%) & 1.42\,(1.3\%) &   10$^8$\\[0.1cm]
$^7$Be & EC & 0.862 (89.5\%) & 4.85\,(7.4\%) & 4.52\,(7.3\%) & 4.88\,(8.1\%) & 10$^9$\\
 & & 0.384 (10.5\%) & &  & \\[0.1cm]
$^8$B & $\beta^+$ & $\lesssim$ 15  & 5.03\,(13\%) & 4.31\,(13\%) & 5.07\,(15\%) & 10$^6$ \\[0.1cm]
hep & $\beta^+$  & $\leq 18.77$ & 7.95\,(31\%) & 8.16\,(31\%) & 7.93\,(30\%) & 10$^3$\\[0.1cm]
$^{13}$N &  $\beta^+$   & $\le$ 1.198 & 2.80\,(16\%) & 2.22\,(13\%) & 3.10\,(15\%) & 10$^8$ \\[0.1cm]
$^{15}$O &  $\beta^+$ & $\le$ 1.732 & 2.07\,(18\%) & 1.58\,(16\%) & 2.30\,(18\%) & 10$^8$ \\[0.1cm]
$^{17}$F &  $\beta^+$  & $\le$  1.738 & 5.35\,(20\%) & 3.40\,(16\%) & 4.70\,(17\%) & 10$^6$ \\[0.1cm] 
\hline 
\end{tabular}
\end{table*}

Solar neutrino cross sections are obtained by integrating over a flux distribution normalized to unity,
denoted below by the probability distribution $P(E_\nu)$
\begin{equation} 
\langle \sigma \rangle = \int^\infty_0 dE_\nu \, P(E_\nu)~ \sigma(E_\nu).
\label{eq:probability}
\end{equation} 
This quantity depends on the shape of the neutrino spectrum for a given source, but not the source's SSM flux.\\
~\\
{\it $\beta$-decay sources:}  The Sun's $\beta^+$-decay sources produce low-energy neutrinos in approximately allowed spectra
\begin{equation}
\begin{split}
\frac{d \omega_{\beta^\pm} }{dE_e} = \frac{G_\mathrm{F}^2 \cos^2 \theta_\mathrm{C}}{2 \pi^3}\, p_e E_e E_\nu^2\, &F_{\beta^\pm} (Z_f,E_e)  \left[ B_{\mathrm{F}^\mp} + g_A^2 B_{\mathrm{GT}^\mp}  \right],
\end{split}
\end{equation}
where $p_e$ and $E_e$ are the electron three-momentum and energy, and $E_\nu$ is the neutrino energy. We adopt the PERKEO3 \cite{PhysRevLett.122.242501} value for the axial coupling constant $g_A=1.2764$ and the Particle Data Group (PDG) values for the Fermi constant $G_\mathrm{F}/(\hbar c)^3=1.1664\times 10^{-5}/$GeV$^2$ and Cabibbo angle $\cos\theta_\mathrm{C}=0.9733$. 
$F_{\beta^\pm} (Z_f,E_e)$ accounts for the distortion of the outgoing positron or electron by the Coulomb field of the daughter nucleus of charge $Z_f$, and
$B_{\mathrm{F}^\mp}$ and $B_{\mathrm{GT}^\mp}$ are the Fermi and Gamow-Teller (GT) transition probabilities, respectively, with the former contributing only to transitions between members of the same isospin multiplet. The superscript $\mp$ on these probabilities indicate whether the nucleus's isospin is being lowered $(p \rightarrow n)$ or raised $(n \rightarrow p)$.

The energy released in the decay is denoted by $W_0$, the nuclear mass difference between the initial and final states,
which up to small atomic binding corrections, is equivalent to $M[i]-M[f]-m_e$, where $M$ is the atomic mass.  The electron and neutrino energies are related by $W_0=E_e+E_\nu$, neglecting tiny corrections due to nuclear recoil and electronic rearrangement. Thus 
\begin{equation}
P (E_\nu,W_0) = N p_e \, E_e  \,E_\nu^2 \, F_{\beta^+}(Z_f,E_e),
\label{eq:naive}
\end{equation}
where the constant $N$ is fixed by the normalization condition
\begin{equation}
    \int_0^{E_\nu^\mathrm{max}}dE_\nu~ P(E_\nu,W_0) \equiv 1.
\end{equation}
The maximum neutrino energy, $E_\nu^\mathrm{max} =W_0-m_e$, is listed in Table \ref{tab:fluxes} for the solar sources of interest.   In the cases described below where $F_{\beta^+}(Z_f,E_e)$ is computed, we follow \cite{PhysRevC.108.035502},
with the electron amplitude at the nuclear surface taken from solutions of the Dirac equation for a uniform
charge distribution, with corrections for
finite nuclear size and a more realistic charge distribution added through the terms $L_0$ and $U$ discussed below. The various $\beta$-decay sources are treated as follows:

$^{13}$N, $^{15}$O, \textit{and} $^{17}$F \textit{neutrinos}:  These spectra are computed from Eq. \eqref{eq:naive}, treating the effects of Coulomb distortion on the outgoing positron as described above. The error incurred by working in the allowed limit can be estimated from the leading correction due to weak magnetism, which is  $O(W_0/m_N)$, or about 0.1\% and thus negligible.

pp \textit{neutrinos}:  Solar reactions occur in a high-temperature plasma.  The effects of temperature
on the $\beta$-decay spectra discussed above, as well as on $^8$B spectrum discussed below, are negligible \cite{PhysRevD.44.1644}: the first-order Doppler shifts due to the motion
of the parent nucleus cancel, as there is no preferred direction.  In contrast, for the $\beta$-decay fusion reaction $p+p$, finite temperature modifies the relative kinetic energy of the fusing ions.  The net effect, after integrating over the thermal profile of the solar core, can be well approximated by introducing a spectrum shift $\Delta W_0^\mathrm{pp} \approx 3.2$  keV  \cite{Bahcall:1997eg},
\begin{equation}
\begin{split}
P_\mathrm{lab}(E_\nu,W_0)  &\rightarrow    P_\odot (E_\nu,W_0^\mathrm{th}) \equiv P_\mathrm{lab} (E_\nu,W_0 + \Delta W_0^\mathrm{pp}).
\end{split}
\label{eq:shift}
\end{equation}
In this case, the shift is a significant  $\approx 1\%$ of $E_\nu^\mathrm{max}$.

$^8$B \textit{neutrinos}:  This spectrum extends to about 17 MeV, and as the decay is dominated by the transition to the broad 2$^+$ resonance at $\approx$ 3 MeV in $^8$Be, $\Gamma_\mathrm{cm} \approx 1.5$ MeV,
deviates significantly from an allowed shape.  As described in the Solar Fusion III (SFIII) study \cite{Acharya:2024lke},  corrections for weak magnetism are significant and are constrained by measurements of the analog electromagnetic decay, while radiative corrections have been evaluated theoretically and are smaller.  We follow SFIII in using the spectrum of Longfellow et al. \cite{PhysRevC.107.L032801}, extracted from the measured spectrum of $\alpha$ particles produced in the resonance decay.  In SFIII, it is noted that
differences in this spectrum relative to the previously adopted standard,  that of Winter et al. \cite{PhysRevC.73.025503}, are less than 5\% for neutrino energies below 15 MeV.  

hep \textit{neutrinos}:  We have taken this spectrum from Bahcall \cite{Bahcall:1997eg}, who assumed an allowed shape. However, as discussed in SFIII, the allowed contribution to the hep reaction is suppressed because the Gamow-Teller operator cannot connect the main $s$-wave components of the initial $p$+$^3$He and final $^4$He states, enhancing the importance of $p$-waves, two-body currents, weak magnetism, and other corrections.  Unfortunately, the hep studies that have incorporated such corrections into cross-section calculations have not published neutrino spectra.  Thus the effects of these corrections on the spectrum shape are presently unknown.  As discussed above for the pp reaction, the hep endpoint will
be shifted by thermal effects, but the shift is  inconsequential ($\approx$ 0.1\% \cite{Bahcall:1997eg}) as $W_0$ is relatively large.

EC \textit{neutrinos}:  The thermal effects include a first-order Doppler broadening of the line due to the motion of the decaying nucleus, and an effective shift associated with the
kinetic energy of the captured electron.  (Continuum electron capture accounts for over 80\% of the total solar capture rate on $^7$Be.)  The $^7$Be line shape has been calculated by Bahcall \cite{PhysRevD.49.3923}.  The result can be expressed in the simple form of Eq. \eqref{eq:shift} with  $\Delta W_0^\mathrm{^7Be}=1.29$ and 1.24 keV for the 862 and 384 keV lines, respectively.  As the detailed profiles are available \cite{PhysRevD.44.1644}, we have instead chosen to numerically integrate over the line shapes.

There does not appear to be a corresponding calculation of the pep line shape in the literature.  However, the effective shift, which includes contributions from both the
electron kinetic energy and the center-of-mass energy of the fusing protons, can be estimated from results in \cite{PhysRevD.44.1644}. Using Eq. (56) of that reference along with the digitized SSM profiles of the BP98 SSM \cite{BP98}, we find an electron contribution to the shift of 1.36 keV.  (The corresponding BP98 calculation
for the $^7$Be shift yields 1.33 keV, in good agreement with the value given above.)
The shift from the relative kinetic energy of the fusing protons can be equated to that for the pp reaction, 3.2 keV. Combining these results, we find that total shift for Eq. \eqref{eq:shift} is $\Delta W_0^\mathrm{pep} \approx 4.5$ keV.

\section{Solar neutrino cross sections}
\begin{figure}
\includegraphics[scale=0.295]{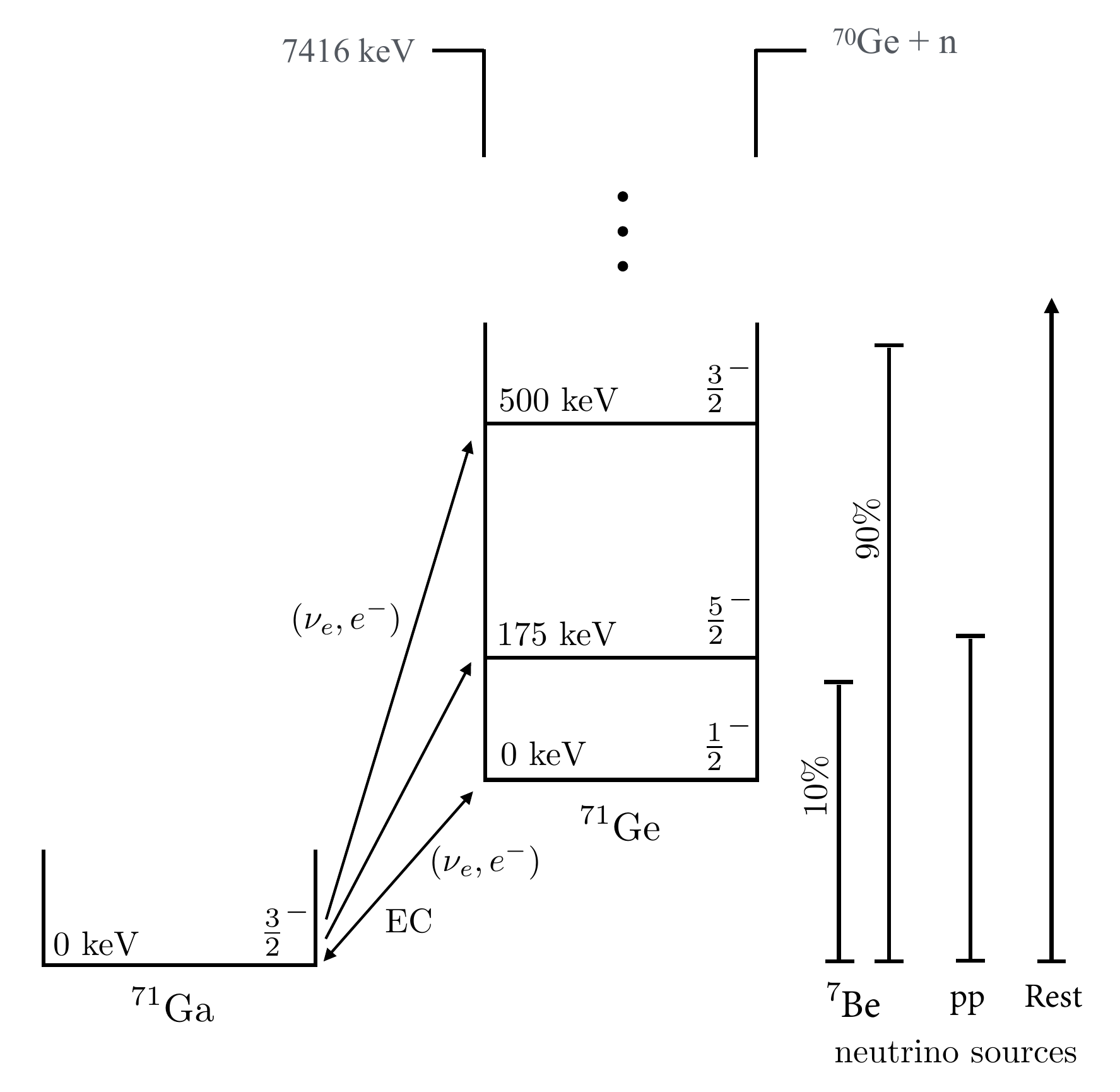}
\caption{Partial level diagram for $^{71}$Ga$(\nu_e,e^-)^{71}$Ge.  The threshold for $^{71}$Ge breakup is 7.416 MeV.}
\label{fig:levels}
\end{figure}

As radiochemical experiments, GALLEX/GNO and SAGE are semi-inclusive, counting transitions to all 
kinematically accessible bound states in $^{71}$Ge --- states below the breakup threshold of 7.416 MeV (see Fig. \ref{fig:levels}).  Near-threshold continuum states with significant $\gamma$-ray branches may also contribute.

The flux-averaged solar neutrino absorption cross section is given in the allowed approximation by
\begin{equation}
    \begin{split}
\langle  \sigma \rangle  = \frac{G_\mathrm{F}^2 \cos^2{\theta_\mathrm{C}} }{  \pi}   \sum_{f} \int& dE_\nu~ P_\odot(E_\nu,W^\mathrm{th}_0)\\
&\times p_e E_e{F}_{\beta^-}(Z_f,E_e)  g_A^2 B_{\mathrm{GT}^+},
 \label{eq:sigmaor}
   \end{split}
 \end{equation}
 where the GT nuclear transition probability for $i \rightarrow f$ is
 \begin{equation}
B_\mathrm{GT^+}=
 \frac{1}{2J_i+1}|\braket{J_f^\pi \alpha_f \|\hat{O}_\mathrm{GT^+}^{J=1} \|J_i^\pi \alpha_i}|^2,
~~~~
    \hat{O}_\mathrm{GT^+}^{J=1}\equiv \sum_{i=1}^A\boldsymbol\sigma(i)\tau_+(i).
    \label{eq:BGT+}
\end{equation}
Here $||$ denotes a matrix element reduced in angular momentum, while $\hat{O}_\mathrm{GT^+}^{J=1}$ is the one-body contribution to the long-wavelength axial-vector current operator.
States are labeled by their spin and parity $J^\pi$, with $\alpha$ representing all other quantum numbers.
The integral in Eq. \eqref{eq:sigmaor} is performed over one of the normalized $P(E_\nu)$ described in the previous section --- either $\beta$-decay spectra or thermally broadened electron-capture line spectra --- while the sum extends over
all kinematically accessible final states.  The atomic mass difference --- the $^{71}$Ge EC Q value --- is precisely known \cite{ALANSSARI20161} 
\begin{equation}
\begin{split}
Q_\mathrm{EC}&=M\left[^{71}\mathrm{Ge}\right]-M\left[^{71}\mathrm{Ga}\right]=232.443\pm 0.093~\mathrm{keV},
\end{split}
\end{equation}
and governs the kinematic relationship
\begin{equation}
E_e=E_\nu-E -Q_\mathrm{EC}+m_e-0.09 \mathrm{~keV},
\end{equation}
where $E$ is the nuclear excitation energy of the state $|J_f^\pi \alpha_f \rangle$, measured relative to the $^{71}$Ge ground state. Here we follow Bahcall \cite{Bahcall:1997eg} by including a very small correction for energy loss due to electron rearrangement. Only neutrinos with $E_\nu > Q_\mathrm{EC}+0.09$ keV $\equiv Q_\mathrm{EC}^\mathrm{eff}$ can produce $^{71}$Ge, setting a lower bound on the integral in Eq. \eqref{eq:sigmaor}.   We omit target recoil effects of order
$O(E_\nu/M_T) \lesssim 10^{-4}$, with $M_T$ being the mass of the daughter nucleus.

${F}(Z_f,E_e)$ is a correction for the Coulomb distortion of the outgoing electron in the field of the daughter nucleus,
\begin{equation}
\begin{split}
&F(Z_f,E_e)=F_0(Z_f,E_e)L_0(Z_f,E_e)U(Z_f,E_e) S(Z_f,E_e) ,
\end{split}
\end{equation}
where the Fermi factor $F_0$ is given by the Dirac solution for a uniform charge distribution, evaluated
at the nuclear surface
\begin{equation}
\begin{split}
F_0(Z_f,E_e)&=4(2p_e R_N)^{2(\gamma-1)}e^{\pi y}\frac{|\Gamma(\gamma+iy)|^2}{[\Gamma(1+2\gamma)]^2},\\
\gamma&\equiv \sqrt{1-(\alpha Z_f)^2},~~~~~y\equiv \alpha Z_f\frac{E_e}{p_e}.
\end{split}
\end{equation}
For the nuclear radius of $^{71}$Ge, we adopt the value $R_N=5.23$ fm, consistent with elastic electron scattering data. The quantities $L_0$ and $U$ are finite-nuclear-size corrections to $F_0$ that, in combination, yield a result similar to what
would be obtained by solving the Dirac equation for a realistic nuclear charge distribution (see \cite{PhysRevC.108.035502, RevModPhys.90.015008}), while $S$ corrects the outgoing electron wave function for the effects of atomic screening.
These are treated as in \cite{PhysRevC.108.035502}.

As the $^{71}$Ga ground state has $J_i^\pi =  \textstyle{\frac{3}{2}}^-$, the allowed final states have $J_f^\pi = \textstyle{\frac{1}{2}}^-$, $\textstyle{\frac{3}{2}}^-,$ or $\textstyle{\frac{5}{2}}^- $.  The final-state sum in Eq. \eqref{eq:sigmaor} extends over all such $^{71}$Ge states below particle breakup with excitation energy $E_\mathrm{ex} \lesssim E_\nu -Q_\mathrm{EC}$ (see Fig. \ref{fig:levels}); we discuss continuum contributions below.  The principal nuclear physics task is to determine the needed $B_\mathrm{GT}$ values. Below, we discuss transitions to the $^{71}$Ge ground state, where the inverse EC rate determines the transition probability, and to excited states, where the $B_\mathrm{GT}$ values are extracted from forward-angle ($p,n$) scattering.

\subsection{Ground-state \texorpdfstring{$B_\mathrm{GT}$}{BGT} value and neutrino capture rate}
An important attribute of $^{71}$Ga$(\nu_e,e^-)^{71}$Ge is its low threshold, which leads to a solar neutrino capture rate dominated by pp neutrinos.  The pp rate is 
determined by the transition to the $^{71}$Ge ground state, the strength of which is rather precisely determined
by the EC half life of $^{71}$Ge,
\begin{equation}
\begin{split}
\tau_{1/2}\left[^{71}\mathrm{Ge}\right]&=\left\{ \begin{array}{cr} 11.43\pm 0.03~\mathrm{d} & \text{\cite{PhysRevC.31.666}}, \\
11.46\pm 0.04~\mathrm{d} & \text{\cite{PhysRevC.108.L021602}}, \\ 11.468\pm 0.008~\mathrm{d} &\text{ \cite{PhysRevC.109.055501}}. \end{array}\right.
\end{split}
\label{eq:half}
\end{equation}
The  weighted mean and standard deviation of the three measurements yields $\tau_{1/2} = 11.465 \pm 0.008$ d and the rate $\omega =(6.997 \pm 0.005) \times 10^{-7}/ \mathrm{s}$.  The uncertainty is less than 0.1\%.  The measurements reported in  \cite{PhysRevC.108.L021602,PhysRevC.109.055501} are recent, announced subsequent to the analysis of \cite{PhysRevC.108.035502}, though
they are in good agreement with the older result \cite{PhysRevC.31.666} employed there.

The procedure for extracting the ground-state $B_\mathrm{GT}$ value from the EC rate is described in \cite{PhysRevC.108.035502}, and need not be repeated here.  We have updated these results to take into account the new EC measurements.  The  $L/K$ and $M/K$ capture ratios are also needed in the analysis of \cite{PhysRevC.108.035502}, which were taken to be 0.117 and 0.019, respectively.  As noted in \cite{PhysRevC.108.L021602}, newer results are available. The measurements of \cite{PhysRevLett.106.131301,PhysRevC.108.035502} can be combined to give the $L/K$ ratio of $0.118\pm0.004$, while \cite{PhysRevLett.116.071301} yields $M/K= 0.019\pm0.003$.  As these newer measurements are consistent with the values used in \cite{PhysRevC.108.035502}, no update is necessary.  One then obtains for the $B_\mathrm{GT^+}$ value for $^{71}$Ga$(\nu,e^-)$ excitation of the $^{71}$Ge ground state, 
\begin{equation}
\begin{split}
{B}_\mathrm{GT^+}^\mathrm{eff} \mathrm{(g.s.)} &\equiv{B}_\mathrm{GT^+} \mathrm{(g.s.)} [1+g_{v,b}]_\mathrm{EC} [1+ \epsilon_q]_\mathrm{EC} \\
&= 0.0863 \pm 0.0013~(95\%~\,\mathrm{C.L.}),
\end{split}
\label{eq:Beff}
\end{equation}
where the two additional terms are the small radiative and weak-magnetism contributions to EC, discussed below and in \cite{PhysRevC.108.035502}.

\subsection{Extracting \texorpdfstring{$B_\mathrm{GT}$}{BGT} values from charge-exchange reactions}
\label{sec:excited_states}
Kuzmin, who proposed the $^{71}$Ga detector, noted the issue of transitions to $^{71}$Ge excited states \cite{Kuzmin:1965zza}.   It is well established empirically, apart from corrections we discuss
below, that $B_\mathrm{GT}$ profiles can be extracted from medium-energy forward-angle $(p,n)$ scattering \cite{TADDEUCCI1987125}. Measurements were made on $^{71}$Ga by Krofcheck et al. \cite{PhysRevLett.55.1051}, using 120 and 200 MeV proton beams. Later, Frekers et al. obtained similar results with ($^3$He,t) \cite{PhysRevC.91.034608,PhysRevC.100.049901}.

While the highest energy solar neutrinos can excite $^{71}$Ge states up to nearly 20 MeV, only those states that decay electromagnetically, rather than by particle breakup, will contribute to $^{71}$Ge counting. This includes all $^{71}$Ge states below neutron breakup threshold, $S_n(0^+)=7.416$ MeV, the minimum energy to produce a free neutron accompanied by $^{70}$Ge in its 0$^+$ ground state.   In Bahcall's treatment of the $^{71}$Ga cross section \cite{Bahcall:1997eg}, it was assumed that these would be the only contributing states. However, GT transitions from the $3/2^-$ ground state of $^{71}$Ga can populate $1/2^-$, $3/2^-$, and $5/2^-$ states in $^{71}$Ge, of which only the $1/2^-$ and $3/2^-$ states can undergo $p$-wave decay to the $0^+$ ground state of $^{70}$Ge. A channel for $p$-wave decay of the $5/2^-$ states opens up at $S_n(2^+)=8.46$ MeV, when the $2^+$ first excited state of $^{70}$Ge can be reached.  Thus one might anticipate additional contributions to the Ga cross section from (technically) unbound states between 7.41 and 8.46 MeV.

Indeed, from the ratios of $\gamma$-decay widths to total widths, $\Gamma_\gamma/\Gamma$, obtained by Ejiri et al. \cite{EJIRI1998257} from $^{71}$Ga($^3$He,t$\gamma$), we find that $58\pm 4$ and $46\pm 4$\% of the transition strengths within the last two energy bins of Table \ref{tab:bgt_krofcheck}, respectively, yield bound $^{71}$Ge. We reduce the Krofcheck $(p,n)$ strengths
in these two bins accordingly. As $\Gamma_\gamma/\Gamma$ drops sharply above 8.46 MeV, we do not consider bins above this threshold. 

We previously noted the generally good track record of forward-angle $(p,n)$ measurements of GT strengths, established
empirically by testing the method on GT transitions of known strength.  See \cite{PhysRevC.108.035502} for
examples and further discussion. 
However, cases are known \cite{PhysRevLett.55.1369} where the proportionality can fail, particularly for weak transitions.  Two such transitions contribute to the $^7$Be neutrino response.  In \cite{PhysRevC.108.035502} it was shown that these discrepancies are removed if a tensor correction is added to the dominant spin $(p,n)$ operator
\begin{equation}
\hat{O}^{J=1}_{({p,n})}(\delta) \equiv \hat{O}^{J=1}_\mathrm{GT^+}+\delta \hat{O}^{J=1}_\mathrm{T^+},
\label{eq:EO}
\end{equation}
where
\begin{equation}
\hat{O}_\mathrm{T^+}^{J=1}=\sqrt{8\pi}\sum_{i=1}^A\left[Y_2(\Omega_i)\otimes\boldsymbol\sigma(i)\right]_{J=1}\tau_+(i),
\end{equation}
with $\delta$ its dimensionless coupling strength.
Thus, the quantity extracted from $(p,n)$ scattering for the transition $i \rightarrow f$ is
\begin{equation}
B_\mathrm{GT}^{(p,n)}=\frac{1}{2J_i+1} |\langle J_f^{\pi} \alpha_f || \hat{O}^{J=1}_{(p,n)}(\delta) || J_i^{\pi} \alpha_i \rangle |^2.
\end{equation}
With this correction, a remarkable improvement was achieved in the correlation between $B_\mathrm{GT}^{(p,n)}$ and $B_\mathrm{GT}$, for measurements over a wide range of transition strengths and nuclear targets, for a fixed $\delta =0.074\pm 0.008  \,  (1 \sigma)$ \cite{PhysRevC.108.035502}.   

The need for such a tensor contribution is not surprising: $(p,n)$ scattering at forward angles and thus low momentum transfers
will select out the longest-range component of the nuclear force, one-pion exchange.  The pion-exchange potential
includes both central and tensor terms. When averaged to an effective one-body operator, the former yields a contribution proportional to $\hat{O}^{J=1}_\mathrm{GT}$, the latter to $\hat{O}^{J=1}_\mathrm{T}$. 

The presence of a tensor contribution in $(p,n)$ scattering reactions was first noted in \cite{PhysRevLett.55.1369}, while
the importance of accounting for this contribution when extracting the $B_\mathrm{GT}$ strengths from forward-angle (p,n) 
measurements was first addressed in \cite{Hata:1995cw,HAXTON1998110}.  The analysis performed in \cite{Hata:1995cw,HAXTON1998110} was recently extended in \cite{PhysRevC.108.035502}, using an improved
selection of data and a more careful uncertainty analysis.  That analysis yielded the value of $\delta$ given above.

\subsection{Excited-state \texorpdfstring{$B_\mathrm{GT}$}{BGT} values: the 175 keV and 500 keV levels}
\label{sec:excited_states1}
Transitions to the $5/2^-$ (175 keV) and $3/2^-$ (500 keV) states in $^{71}$Ge are of special importance as they govern the excited-state contributions to the $^{51}$Cr and $^{37}$Ar neutrino-source experiments as well as to pp and $^7$Be solar neutrino capture (see Fig. \ref{fig:levels}).  The 
$(p,n)$ measurements made on $^{71}$Ga \cite{PhysRevLett.55.1051} had sufficient resolution to separate
the contributions of these two states from each other and from the ground state.  The recent calculations of \cite{PhysRevC.108.035502} used the effective operator formalism described above to extract the underlying $B_\mathrm{GT}$ values from the measured $B_\mathrm{GT}^{(p,n)}$ values, propagating all errors via Monte Carlo.  The results are
\begin{equation}
\begin{split}
B_\mathrm{GT^+}(5/2^-) &\le 0.0077 ~ (68\% \, \mathrm{C.L.}),\\
B_\mathrm{GT^+}(3/2^-) &= 0.0104 \pm 0.0022 ~(68\%  \,\mathrm{C.L.}).
\label{eq:resolved_BGTs}
\end{split}
\end{equation}
These extractions depend on nuclear shell-model (SM) estimates of the relative strengths of the spin and spin-tensor operators. The uncertainties above account for the theory differences that result from the use of different candidate effective interactions.  As SM calculations make certain common model assumptions, such variations do not necessarily reflect the entire theory uncertainty.

\subsection{Other excited-state contributions}
\label{sec:excited_states2}  
At energies above those of the two states just considered, the density of $^{71}$Ge states grows rapidly. While one can no longer extract the strengths of individual transitions due to the limited resolution of $(p,n)$ measurements, the broad profile of $B_\mathrm{GT}^{(p,n)}$ can be determined, with transitions grouped into energy bins whose widths are comparable to the resolution.
The bins employed in Table \ref{tab:bgt_krofcheck} are from \cite{PhysRevLett.55.1051,Krof87},
apart from the three highest in energy.  In those cases --- $7.0$--$7.42$, $7.42$--$8.0$, and $8.0$--$8.46$ MeV ---
the bin boundaries have been slightly adjusted so that they correspond to the neutron breakup thresholds for populating the 0$^+$ ground state and 2$^+$ first excited state of ${}^{70}$Ge.  In these bins, we have
then reduced the $B^{(p,n)}_\mathrm{GT}$ strength, retaining the proportion that decays electromagnetically and
thus contributes to $^{71}$Ge counted in the SAGE and GALLEX/GNO experiments.  We take the
$\gamma$-decay branching probability from \cite{EJIRI1998257}.

\begin{table*}
\centering
\caption{GT strengths $B_\mathrm{GT}^{(p,n)}$ extracted from $(p,n)$ reactions \cite{PhysRevLett.55.1051,Krof87}.  Except for the first three transitions, the strengths are summed over all states within an energy bin.  The bins are taken from \cite{Krof87} except for the last three, where some slight adjustments have been made to align with neutron breakup thresholds (see text).  The derived $B_\mathrm{GT}$ values take into account both the tensor contribution to $B_\mathrm{GT}^{(p,n)}$ and, for the last two continuum bins, the fraction of the strength $\Gamma_\gamma/\Gamma$ that decays electromagnetically and thus contributes to $^{71}$Ge counting \cite{EJIRI1998257}. 
}
\label{tab:bgt_krofcheck}
\begin{threeparttable}
{\renewcommand{\arraystretch}{1.1}
\begin{tabular}{lcccclccc}
\hline
\hline
$E$ (MeV) & $B_\mathrm{GT}^{(p,n)}$ & $\Gamma_\gamma/\Gamma$ & $B_\mathrm{GT^+}$ &~~~~~ & $E$ (MeV) & $B_\mathrm{GT}^{(p,n)}$ & $\Gamma_\gamma/\Gamma$ & $B_\mathrm{GT^+}$\\
\hline
0.0 & $0.089 \pm 0.007$ & -- & $0.0862 \pm  0.0012$\tnote{\hyperref[tnote1]{a}} & & $4.0$--$4.5$ & $0.2937 \pm 0.0213$ & 1 & $ 0.2972 \pm 0.0225 $\\
0.175 & $<0.005$ & 1 & $\le 0.0077$ & & $4.5$--$5.0$ & $0.3008 \pm 0.0210$ & 1 & $ 0.3003 \pm 0.0214 $\\
0.500 & $0.011\pm 0.002$ & 1 & $0.0104 \pm 0.0022$ & & $5.0$--$5.5$ & $0.2906 \pm 0.0206$ & 1 & $ 0.2827 \pm 0.0205 $\\
$0.6$--$1.0$ & $0.0401 \pm 0.0039$ & 1 & $ 0.0405\pm 0.0040 $ & & $5.5$--$6.0$ & $0.4794 \pm 0.0319$ & 1 & $ 0.4508 \pm 0.0324 $\\
$1.0$--$1.5$ & $0.0360 \pm 0.0037$ & 1 & $ 0.0368 \pm 0.0038 $  & & $6.0$--$6.5$ & $0.5292 \pm 0.0349$ & 1 & $ 0.4783 \pm 0.0368 $\\
$1.5$--$2.0$ & $0.0432 \pm 0.0045$ & 1 & $ 0.0437 \pm 0.0046 $ & & $6.5$--$7.0$ & $0.4382 \pm 0.0281$ & 1 & $ 0.3856 \pm 0.0295 $\\
$2.0$--$2.5$ & $0.0394 \pm 0.0038$ & 1 & $ 0.0395 \pm 0.0039 $ & & $7.0$--$7.42$ & $0.3920 \pm 0.0254$ & 1 & $ 0.3428 \pm 0.0249 $\\
$2.5$--$3.0$ & $0.0980 \pm 0.0089$ & 1 & $ 0.0993 \pm 0.0092 $ & & $7.42$--$8.0$ & $0.5097 \pm 0.0287$ & $0.58\pm 0.04$ & $ 0.2604 \pm 0.0240 $\\
$3.0$--$3.5$ & $0.1495 \pm 0.0128$ & 1 & $ 0.1516 \pm 0.0130 $ & & $8.0$--$8.46$ & $0.2935 \pm 0.0185$ & $0.46\pm 0.04$ & $ 0.1208 \pm 0.0130 $\\
$3.5$--$4.0$ & $0.2444 \pm 0.0189$ & 1 & $ 0.2463 \pm 0.0200 $ & & & &\\
\hline
\hline
\end{tabular}
}
\begin{tablenotes}
    \item[a] \label{tnote1} \small{Ground-state GT strength, taken from EC rate, corresponds to $B_\mathrm{GT^+}^\mathrm{eff}$ defined in Eq. \eqref{eq:Beff}.}
\end{tablenotes}
\end{threeparttable}
\end{table*} 

From the tabulated $B_\mathrm{GT}^{(p,n)}$ strength distribution, one can extract the $B_\mathrm{GT^+}$ 
distribution needed in calculations of the $^8$B and hep neutrino cross sections on $^{71}$Ga, provided the data are corrected for the tensor contribution.  While this correction has been made previously for the resolved states at 175 and 500 keV in $^{71}$Ge, 
we are not aware of any attempt to evaluate the impact of the tensor correction on the inclusive GT response.

The standard SM space for $A$=71 nuclei is $2p_{3/2}1f_{5/2}2p_{1/2}1g_{9/2}$: several
carefully tuned, phenomenologically successful effective interactions exist for this space, and the space
is separable, so that spurious center-of-mass excitations can be eliminated.  Inclusive response functions can in principle
be evaluated by solving for a complete set of $^{71}$Ge states, then summing transitions to those states.
In practice, this is impractical as the dimension $N$ of the $m$-scheme SM basis 
for $^{71}$Ge exceeds $10^8$.  Here we use two powerful and complementary techniques
to characterize the inclusive response function,
\begin{enumerate}
\item Energy-unweighted sum rules, computed from the SM $^{71}$Ga ground state: these give
information about the total strength, but not its spectral distribution;
\item An iterative Lanczos moments method \cite{ChalkRiver,Lanczos} that
extracts spectral information from the SM Hamiltonian $\hat{H}_\mathrm{SM}$, though our preferred SM
space is incomplete.
\end{enumerate}

\noindent
{\it Sum rules:}  The vector operators of interest
\begin{equation}
    \begin{split}
 \hat{O}^{1M}_{(p,n)}(\delta) &=   \sum_{i=1}^A \left[ \sigma_{1M} (i) + \delta \sqrt{8\pi} \left[Y_2(\Omega_i)\otimes \boldsymbol{\sigma}(i)\right]_{1M} \right] \tau_+(i), \nonumber \\
  \hat{O}^{1M}_\mathrm{GT^+}  &=  \sum_{i=1}^A \sigma_{1M} (i)\tau_+(i),
  \end{split}
  \label{eq:Lops}
\end{equation}
act on the SM $^{71}$Ga ground state to form a vector $|w^{1M} \rangle$,
\begin{equation}
    \hat{O}^{1M} \,|^{71}\mathrm{Ga}(\mathrm{g.s.}) \rangle = |w^{1M} \rangle.
\end{equation}
While the ground state resides in the $2p_{3/2}1f_{5/2}2p_{1/2}1g_{9/2}$ valence space, $|w \rangle$ includes components outside this space. For example, $\hat{O}^{1M}_\mathrm{GT}$ connects the orbitals  $1f_{5/2} \leftrightarrow1f_{7/2}$, and $1g_{9/2} \leftrightarrow 1g_{7/2}$. One then uses closure to compute
the norm of  $|w^{1M} \rangle$, with holes below and particles above the valence space contracting.
This requires evaluation of the ground-state one- and two-body density matrices.  One obtains, for example,
 \begin{equation} 
\frac{1}{2J_{i}+1} \sum_{J^\pi_f \alpha_f} \left| \langle J^\pi_f \alpha_f || \hat{O}^1_\mathrm{GT^+} || J^\pi_{i} \alpha_{i}  \rangle \right|^2 = \sum_{M=-1}^1 \langle w^{1M}_\mathrm{GT^+} | w^{1M}_\mathrm{GT^+} \rangle ,
 \end{equation}
giving the familiar Ikeda rule discussed below.  The sum rule includes all excitations and thus yields
the correct integrated strength of the response function, but does not provide spectral information.\\

\noindent
{\it Lanczos method:} One operates with $\hat{O}^{1M}$ on the SM $^{71}$Ga ground state to form a vector $|w^{1M} \rangle_R$
\begin{equation}
 \hat{O}^{1M}  \left|^{71}\mathrm{Ga}(\mathrm{g.s.}) \right\rangle =  |w^{1M} \rangle_R \equiv N_R^{1M} \,|v^{1M} \rangle_R,\\
 \label{eq:Lanczos1}
\end{equation}
where $|v^{1M}\rangle_R$ is a unit vector, $N^{1M}_R$ is a normalization, and the subscript $R$ indicates that only 
configurations within the SM space are retained.  We take $|v^{1M}\rangle_R \equiv |v_1\rangle$ as the pivot for recursively constructing the Lanczos matrix 
\begin{equation}
    \begin{split}
\hat{H}_\mathrm{SM} |v_1 \rangle &= \alpha_1 |v_1 \rangle + \beta_1 |v_2 \rangle, \nonumber \\
\hat{H}_\mathrm{SM} |v_2 \rangle &= \beta_1 |v_1 \rangle + \alpha_2 |v_2 \rangle + \beta_2 |v_3 \rangle, \nonumber \\
\hat{H}_\mathrm{SM} |v_3 \rangle &= \hspace{1.3cm} \beta_2 |v_2 \rangle + \alpha_3 |v_3 \rangle + \beta_3 |v_4 \rangle, ~~~\ldots
 \end{split}
\end{equation}
where $\hat{H}_\mathrm{SM}$ is the SM Hamiltonian.  As $\hat{H}_\mathrm{SM}$ is only defined within the
valence space, all vectors $|v_i \rangle$ are contained in that space.

One truncates the algorithm after $n$ steps, where $n \ll N$, with $N$ being the dimension of the SM basis.  The Lanczos matrix is then diagonalized, yielding a set of $n$ energies and eigenstates $\left\{E_L(i) , |\phi_L(i) \rangle, \, i=1,n \right\}$.  Let us denote the exact eigenvalues and eigenenergies --- the results one would obtain if one could diagonalize the full $\hat{H}_\mathrm{SM}$ of dimension $N$ --- as $\left\{ E(i) , |\phi(i) \rangle , \, i=1,N \right\}$. One can demonstrate that
 \begin{equation}
 \begin{split}
 \langle v_1 | \hat{H}_\mathrm{SM}^\lambda |v_1 \rangle &=  \sum_{i=1}^n \left|  \langle v_1 | \phi_L(i) \rangle \right|^2 E^\lambda_L(i) \equiv \sum_{i=1}^n f_L(i)  E^\lambda_L(i) \\
 &= \sum_{i=1}^N \left|  \langle v_1 | \phi(i) \rangle \right|^2 E^\lambda(i),
 \end{split}
 \end{equation}
 for $\lambda=1, \ldots, 2n-1$. That is, the $n$ Lanczos energies $E_L(i)$ and associated weights $ f_L(i) \equiv \left|  \langle v_1 | \phi_L(i) \rangle \right|^2$ and the $N$ exact SM energies $E(i)$ and weights $\left| \langle v_1 | \phi(i) \rangle \right|^2$  form two discrete probability
 distributions whose $2n-1$ lowest moments in energy are identical.  The Lanczos algorithm extracts from the full $N$-dimensional $\hat{H}_\mathrm{SM}$ exactly the long-wavelength  moments needed to characterize the broad profile of the response function.  If one is comparing theory to an experiment with limited resolution,
 so that the experiment is insensitive to the high-frequency variations of the response, then the Lanczos procedure provides precisely the information needed. While the Lanczos representation of the response function is discrete --- a set of energy points and weights --- a continuous distribution can be obtained by spreading the $\delta$-function strengths $f_L(i)$ around the points $E_L(i)$ using a smoothing function, such as a Gaussian or Lorentzian. We adopt the latter choice, leading to the replacement 
 \begin{equation}
 \begin{split}
 &\sum_{i=1}^n \int dE \, f_L(i) \, \delta(E-E_L(i)) \\
 &\rightarrow \sum_{i=1}^n \int dE \, f_L(i) \, \frac{\Gamma}{2 \pi} \frac{1}{
 (E-E_L(i))^2 + (\Gamma/2)^2}    \equiv \int dE \,\tilde{f}_n(E).
  \end{split}
 \end{equation}
 The Lorentzian width $\Gamma$ can be chosen to mimic the resolution of the comparison experiment.  One
 expects the continuous function $\tilde{f}_n(E)$ to converge with increasing $n$, as the typical spacing of neighboring Lanczos eigenvalues 
 becomes small compared to $\Gamma$. This smoothing allows one to more easily calculate the effects of the tensor correction as a function of $E$ --- that is,
with fewer Lanczos iterations $n$ --- since it damps the fluctuations that one would encounter when using the corresponding discrete probability distribution.  The correction one
 applies to extract $B_\mathrm{GT}$ from the experimental results for $B_\mathrm{GT}^{(p,n)}$ is 
 \begin{equation}
 \Delta(\epsilon_i,\epsilon_f)\equiv \frac{  \displaystyle{\sum_{M=-1}^1}  ( N_{R,\mathrm{GT}}^{1M})^2 \int_{\epsilon_i}^{\epsilon_f} \, \tilde{f}^{1M}_\mathrm{GT}(E) \,dE }{  \displaystyle{\sum_{M=-1}^1}  ( N_{R,\mathrm{GT+ T}}^{1M})^2 \int_{\epsilon_i}^{\epsilon_f} \, \tilde{f}^{1M}_\mathrm{GT+ T}(E) \, dE },
 \label{eq:ratio}
  \end{equation}
 for each energy bin $[\epsilon_i,\epsilon_f]$ in Table \ref{tab:bgt_krofcheck}.  The sum extends over the three magnetic
 projections of the operators defined in Eq. \eqref{eq:Lops}.  Note that the correction factor involves a matrix element ratio; this reduces worries about the effects of configurations outside the SM space.  The resulting corrected $B_\mathrm{GT}$ values are shown in Table \ref{tab:bgt_krofcheck}. The error bars include the nuclear modeling uncertainties from the three SM calculations described below.

\begin{table}
\centering
\caption{The ratio of single-particle matrix elements $\langle f\|O_\mathrm{T}^{J=1}\|i\rangle/\langle f\|O_\mathrm{GT}^{J=1}\|i\rangle$.} 
\label{tab:M_ratios}
{\renewcommand{\arraystretch}{1.5}
\begin{tabular}{llr}
\hline
\hline
$\langle f\|$ & ~~~~$\|i\rangle$ & Ratio\\
\hline
$\langle\left(\ell\frac{1}{2}\right)j=\ell-\frac{1}{2}\|$ & ~~~~$\|\left(\ell\frac{1}{2}\right)j=\ell-\frac{1}{2}\rangle$ & $\frac{2(\ell+1)}{2\ell-1}$\\
$\langle\left(\ell\frac{1}{2}\right)j=\ell+\frac{1}{2}\|$ & ~~~~$\|\left(\ell\frac{1}{2}\right)j=\ell+\frac{1}{2}\rangle$ & $\frac{2\ell}{2\ell+3}$\\
$\langle\left(\ell\frac{1}{2}\right)j=\ell-\frac{1}{2}\|$ & ~~~~$\|\left(\ell\frac{1}{2}\right)j=\ell+\frac{1}{2}\rangle$ & $-\frac{1}{2}$\\
 $\langle\left(\ell\frac{1}{2}\right)j=\ell+ \frac{1}{2}\|$ & ~~~~$\|\left(\ell\frac{1}{2}\right)j=\ell-\frac{1}{2}\rangle$& $-\frac{1}{2}$\\
$\langle\left(\ell\frac{1}{2}\right)j=\ell+\frac{1}{2}\|$ & ~~~~$\|\left((\ell+2)\frac{1}{2}\right)j=\ell+\frac{3}{2}\rangle$ & $\pm \infty$\\
\hline
\hline
\end{tabular}
}
\end{table}

\subsection{Sum rule results}
Table \ref{tab:M_ratios} gives the ratio of the tensor and GT amplitudes for single-particle matrix elements.
The interference between the operators can be constructive or destructive for spin-orbit partners, depending on whether the
matrix elements are diagonal (e.g., $2p_{3/2} \leftrightarrow 2p_{3/2}$) or not (e.g., $2p_{3/2} \leftrightarrow 2p_{1/2}$). 
Consequently, the shell structure of the ground state plays a role in determining the character of the interference in a given region of the response spectrum. For example, in $^{71}$Ga the amplitude $2p_{1/2}(n) \rightarrow 2p_{3/2}(p)$ is expected to couple the neutron and proton Fermi surfaces; the resulting destructive interference would cause $B_\mathrm{GT}^{(p,n)}$ to underestimate $B_\mathrm{GT^+}$ at very low excitation energies.
    
The best known sum-rule result for GT responses is that of Ikeda \cite{10.1143/PTP.31.434,FUJITA1965145}, which constrains the difference between $B_\mathrm{GT^+}$ and $B_\mathrm{GT^-}$, where the latter governs the response for the conjugate reaction $^{71}$Ga$(\bar{\nu}_e,e^+)^{71}$Zn. ($B_\mathrm{GT^-}$ is defined in analogy with Eq. \eqref{eq:BGT+}, but with the isospin operator change $\tau_+ \rightarrow \tau_-$.  The same
substitution distinguishes $B_\mathrm{GT}^{(n,p)}$ from $B_\mathrm{GT}^{(p,n)}$.) It is straightforward to generalize the Ikeda sum rule to charge exchange, 
\begin{equation}
\ \sum_{J^\pi_f \alpha_f} \left[ B_\mathrm{GT}^{(p,n)}-B_\mathrm{GT}^{(n,p)} \right]  \,=\, 3(1+2 \delta^2) (N-Z).
\end{equation}
(The standard Ikeda result is obtained by taking $\delta \rightarrow 0$, as this removes the tensor operator.)
The absence of a term linear in $\delta$ ensures that the total strength in the difference between the $(p,n)$ and $(n,p)$ 
responses will deviate from the weak-interaction Ikeda sum rule by only $\approx$ 1\%.  It also provides a first hint that
effects linear in $\delta$ primarily redistribute strength, while leaving the integrated strength largely unchanged.

\begin{table*}
\centering
{\renewcommand{\arraystretch}{1.2}
\begin{tabular}{ccccc}
\hline
\hline
 g.s. description & restrictions & ~~~~~~~~$\sum B_\mathrm{GT}^{(p,n)}$~~~~~~~~ & ~~ ~~~~~~$\sum B_\mathrm{GT}^{(n,p)}$~~~~~~~~ & ~~~$\sum [B_\mathrm{GT}^{(p,n)}-B_\mathrm{GT}^{(n,p)} ]$~~~\\
 \hline
  $2p_{3/2}1f_{5/2}2p_{1/2}1g_{9/2}$ &  none  & $29.55-2.58\delta+156.45\delta^2$ & $2.55-2.58\delta+102.45\delta^2$ & $27(1+2\delta^2)$  \\
 " &  SM  & $13.23+13.73\delta+26.30\delta^2$ & $0.31-0.34\delta+0.32\delta^2$ & $12.92+14.07\delta+25.98\delta^2$ \\
 $1f_{7/2} 2p_{3/2}1f_{5/2}2p_{1/2}$ & none & $27+153.86\delta^2$ & $99.86\delta^2$ & $27(1+2\delta^2)$ \\
\hline
\hline
\end{tabular}
}
\caption{Energy-independent sum rules for $B_\mathrm{GT}^{(p,n)}$, $B_\mathrm{GT}^{(n,p)}$, calculated for a complete set of states
via the ground-state one- and two-body density matrices, or alternatively, calculated with only those final states contained within the
indicated SM space.  If $\delta= 0$, the corresponding results for $B_\mathrm{GT^+}$ and $B_\mathrm{GT^-}$ are obtained.
The ground state is computed either in the canonical $2p_{3/2}1f_{5/2}2p_{1/2}1g_{9/2}$ space or in $1f_{7/2} 2p_{3/2}1f_{5/2}2p_{1/2}$,
a space that includes all spin partners but requires us to describe $^{71}$Ga with a closed neutron shell.  These results were generated from the JUN45 and KB3G interactions, respectively.}
\label{tab:sumrule}
\end{table*}

One can also evaluate the sum rules in the two isospin directions separately, from the ground-state one- and two-body density matrices, e.g.,
\begin{equation}
  \sum_{f}B_\mathrm{GT^\pm}\equiv\frac{1}{2J_i+1}\sum_f |\braket{J_f\alpha_f\|\hat{O}_\mathrm{GT^\pm}^{J=1}\|J_i\alpha_i}|^2,
\label{eq:Ikedag}
\end{equation}
and similarly for $B_\mathrm{GT}^{(p,n)}$ and $B_\mathrm{GT}^{(n,p)}$.  Results for the latter, where the $^{71}$Ga ground state is described within the SM space $2p_{3/2}1f_{5/2}2p_{3/2}1g_{9/2}$ but transitions to a complete set of final states are summed, are given in the first line of Table \ref{tab:sumrule}.  As our generalized Ikeda sum rule requires, the contribution proportional to $\delta$ is identical in the
two isospin directions.  The linear term is quite small, again hinting that the primary effect of the GT-tensor interference is a
redistribution of strength in each isospin direction, while the integrated strength remains little changed. 

We then repeat this calculation, limiting the sum over final states to those in our SM space. (This requires
removing from the sum rule terms where particles and holes outside the SM valence space contract.)  This incomplete sum
should roughly correspond to an integration over the low-energy spectrum, as these are the states the SM describes.  The
results are given in the second line of Table \ref{tab:sumrule}.  We now see that 1) terms linear in $\delta$ dominate and
constructively interfere, and 2) the $(p,n)$ response exceeds the GT$^+$ response by a fractional amount of $O(\delta) \approx$ 10\%,
which is significant.  The SM states exhaust somewhat less than 50\% of Ikeda sum rule: this suggests that states
outside the SM valence space, 
which should dominate the high-energy side of the GT response, will exhibit destructive interference also roughly proportional to $\delta$.

The tentative conclusion is that the primary effect of the tensor term --- which generates the difference between $B_\mathrm{GT^+}$ and $B_\mathrm{GT}^{(p,n)}$ ---
is a spectral distortion linear in $\delta$, while the overall change in the energy-independent sum rule is much smaller.  We can now use the Lanczos moments method to extract more information on the spectral distortion.

\subsection{Lanczos method results}
As in the last sum-rule example discussed above, this method addresses the impact of the tensor interaction on
the low-energy spectrum, the $\approx$ 50\% of the response carried by states in the SM valence space.
This is the portion of the spectrum, below particle breakup, that dominates solar neutrino capture.  The Lanczos
method goes well beyond the sum rule by determining the shape of the response function, making use of
the spectral information encoded in the effective interaction $\hat{H}_\mathrm{SM}$.
The Lanczos iterations were performed using the configuration-interaction code BIGSTICK \cite{Johnson:2013bna,Johnson:2018hrx} for three $2p_{3/2}1f_{5/2}2p_{1/2}1g_{9/2}$ interactions, JUN45 \cite{jun45}, jj44b \cite{jj44b}, and GCN2850 \cite{gcn2850}. These interactions have been used previously in calculations of electroweak observables and yield satisfactory descriptions of the low-lying spectra for the two nuclei of interest. See Ref. \cite{PhysRevC.108.035502} and references therein for details.

\begin{figure*}
\centering
\includegraphics[scale=0.5]{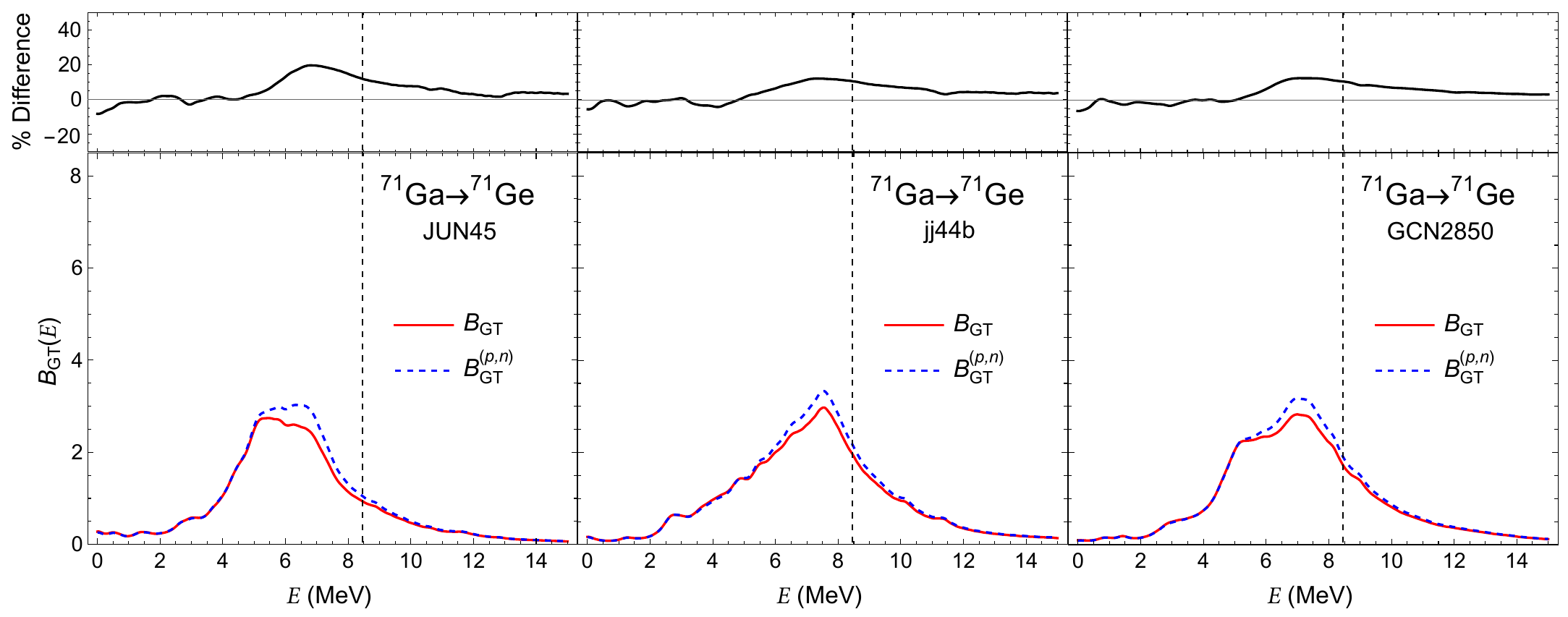}
\caption{Bottom: Response functions for the GT$^+$ and $(p,n)$ operators, calculated by the Lanczos moments method.  The SM basis restricts transitions to the $2p_{3/2}1f_{5/2}2p_{1/2}1g_{9/2}$ valence space. $E$ is the excitation energy of the final state in $^{71}$Ge.  The panels compare results for three commonly used SM effective interactions.  A Lorentzian smoothing function of width $\Gamma = 0.5$ MeV was applied. The vertical dashed line marks the energy $S_{n}(2^+)=8.46$ MeV, above which the branching ratio to the ground state of $^{71}$Ge is effectively zero. Top: Relative difference $[B_\mathrm{GT}^{(p,n)}-B_\mathrm{GT^+}]/B_\mathrm{GT^+}$ due to the inclusion of the tensor operator.}
\label{fig:response_functions_fpg}
\end{figure*}

Figure \ref{fig:response_functions_fpg} shows the GT and $(p,n)$ response functions obtained by performing $n=300$ Lanczos iterations for each of the six needed operators $\hat{O}_\mathrm{GT^+}^{1M}$  and $\hat{O}^{1M}_{(p,n)}$ and for each of the three effective interactions. The differences between the panels provide a measure of the uncertainty stemming from the effective interaction. (This should not be misconstrued as a total theory uncertainty, as the calculations share certain common assumptions.) We chose a smoothing width $\Gamma=0.5$ MeV to approximate the experimental resolution of the $(p,n)$ data. Although there are detailed differences that depend on the choice of interaction, the qualitative changes due to the tensor interaction are very similar.  For low-lying nuclear states with excitation energies $E\lesssim 4$ MeV, the GT and tensor components of $\hat{O}_{(p,n)}$ interfere destructively, so that when one makes the $(p,n)$ tensor correction $\Delta(\epsilon_i,\epsilon_f)$ averaged over the three shell-model calculations, the extracted $B_\mathrm{GT^+}$ strength is \textit{increased} by $\approx 1$--$2\%$ relative to the measured $B_\mathrm{GT}^{(p,n)}$. Above this energy, the interference is constructive, so that when one makes the tensor correction, the extracted $B_\mathrm{GT}$  strength is \textit{reduced} relative to the measured $B_\mathrm{GT}^{(p,n)}$, with the effects growing to $\approx 10\%$ near the peak of the SM response.

The tensor contribution to forward-angle $(p,n)$ scattering was identified from weak GT transitions, where the {\it relative} contribution from $\hat{O}^{J=1}_\mathrm{T^+}$ is enhanced \cite{Hata:1995cw,HAXTON1998110}. The prototypical example is an $\ell$-forbidden M1 transition between orbitals with quantum numbers $[n,\ell,j=\ell+\frac{1}{2}]$ and $[n-1,\ell+2,j=\ell+\frac{3}{2}]$. The Gamow-Teller operator does not change $\ell$, and consequently these transitions will be dominated by the tensor operator. 
Neutrino capture to the first excited state, $^{71}$Ga($3/2^-$, gs)$\rightarrow^{71}$Ge($5/2^-$, $175$ keV), is a candidate $\ell$-forbidden transition: In the na\"ive single-particle SM, a $1f_{5/2}$ neutron is converted to a $2p_{3/2}$ proton. More sophisticated, correlated SM calculations confirm that the density matrix governing this transition does have a strong $\ell$-forbidden component \cite{PhysRevC.108.035502,Kostensalo:2019vmv}.

However, a semi-inclusive response function is an integrated quantity, where ${\it absolute}$ contributions are what matter. As the contribution linear in $\delta$ 
arises from an interference between $\hat{O}_\mathrm{T^+}^{J=1}$ and $\hat{O}_\mathrm{GT^+}^{J=1}$, the tensor operator can make its largest absolute contribution
when the GT amplitude is large.  Numerically, this is what we observe: the largest absolute corrections appear near the peak of the GT response. One can also anticipate the sign of the interference, whether constructive or destructive, based on simple SM arguments.
For example, as noted previously, one would expect the amplitude for conversion of a $2p_{1/2}$ neutron to a $2p_{3/2}$ proton to influence $(\nu,e^-)$ transitions to low-energy states in $^{71}$Ge, as these orbitals are near the neutron and proton Fermi surfaces, respectively.  
From the ratio of single-particle matrix elements of the tensor and GT operators given in Table \ref{tab:M_ratios}, one expects destructive interference and resulting differences between the $(p,n)$ and GT$^+$ responses of $O(\delta) \approx$ 10\%.  This
is the behavior seen numerically below $\approx$ 4 MeV.  As the $2p_{3/2}$ has a significant occupation, Pauli blocking
will tend to suppress the response.

Similarly, one expects the amplitudes $2p_{1/2} \, (n)  \rightarrow 2p_{1/2} \, (p)$ and $1f_{5/2} \, (n)  \rightarrow 1f_{5/2} \, (p)$ to be prominent
in the transition densities for higher-lying levels.  From Table \ref{tab:M_ratios}, one anticipates constructive interference, little Pauli blocking, and thus a strong response.  The absolute contributions should also
correlate with neutron occupations, which would favor the $1f_{5/2}$ orbital over the $2p_{1/2}$. 

\subsection{A SM crosscheck}
There is a nice crosscheck possible in which the Lanczos method is employed in a SM space that includes all spin-orbit partners,
and thus is complete for the GT operator. $^{71}$Ga can be described alternatively as a closed-neutron-shell nucleus in the $1f_{7/2}2p_{3/2}1f_{5/2}2p_{1/2}$ SM space. This description of the $^{71}$Ga ground state is not optimal: $^{71}$Ga is in a transition
region where significant nuclear deformation arises as the neutron number is increased, with polarization effects generating substantial neutron occupation of the $1g_{9/2}$ orbital and proton occupation of the $1f_{5/2}$ orbital, driven in part by the attraction between these two shells \cite{ChalkRiver}.  Yet despite this shortcoming, this space gives us an opportunity
to use the Lanczos method while simultaneously preserving the sum rules for the GT and GT-tensor interference amplitudes. The sum rule results, given in the third line of Table \ref{tab:sumrule}, show that the GT-tensor interference term 
vanishes identically in both isospin directions.  Thus this term can only generate a redistribution of strength.

We then utilize the Lanczos procedure within the  $1f_{7/2}2p_{3/2}1f_{5/2}2p_{1/2}$ SM space, for each of three
effective interactions. As expected, the resulting descriptions of the $^{71}$Ga ground state show deficiencies.  The KB$^\prime$ \cite{kbp} and GXPF1 \cite{gxpf1} interactions incorrectly predict the ordering of the lowest 3 states in $^{71}$Ge. The third interaction, KB3G \cite{kb3g}, yields the correct ordering but gives rather poor values for the energies, placing the $5/2^-$ state at $0.645$ MeV and the $3/2^-$ state at $1.478$ MeV.

\begin{figure*}
\centering
\includegraphics[scale=0.5]{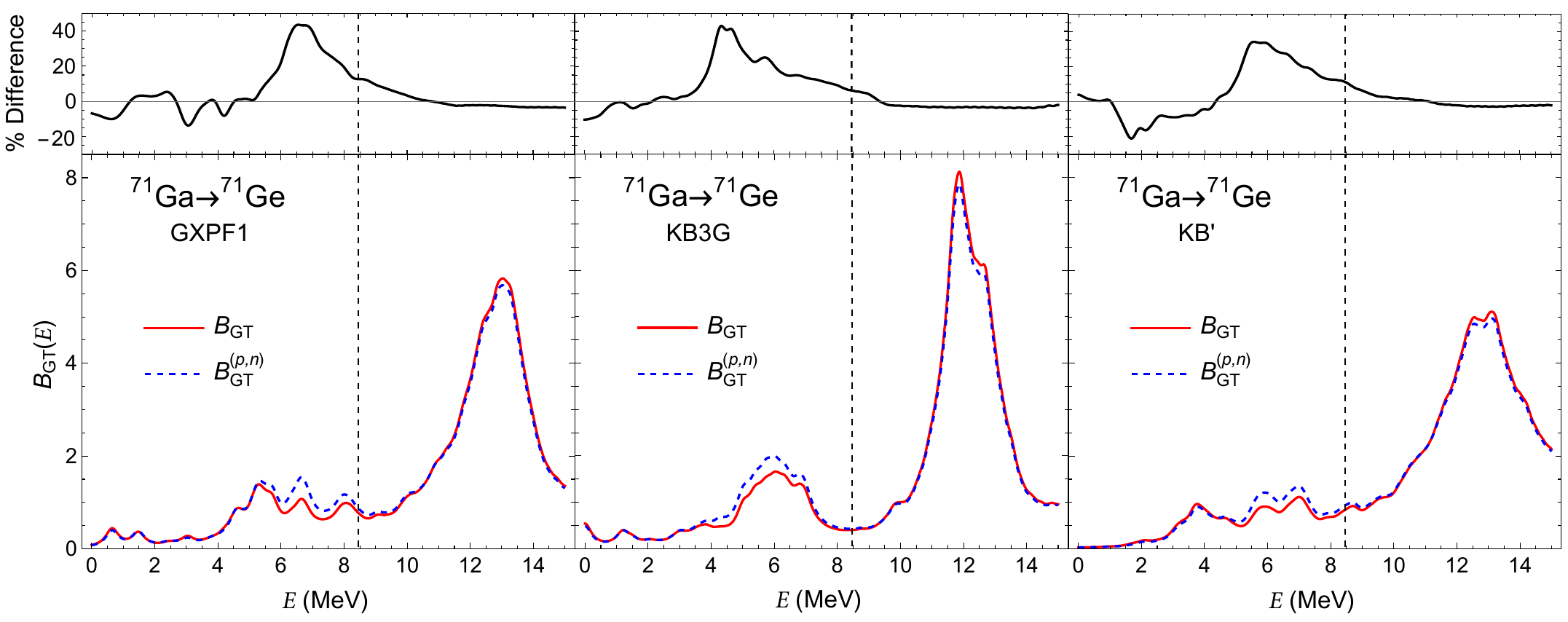}
\caption{As in Fig. \ref{fig:response_functions_fpg} but using 3 different SM interactions in the $1f_{7/2}2p_{3/2}1f_{5/2}2p_{1/2}$ model space.}
\label{fig:response_functions_fp}
\end{figure*}
 
Nevertheless, treating $^{71}$Ga in this space allows us to make several significant points.  First, because the neutron shell is closed, $B_\mathrm{GT^-}$ vanishes and the GT Ikeda sum rule is saturated by $B_\mathrm{GT^+}$.  Consequently, as we see from Table \ref{tab:sumrule}, when the tensor contribution is turned on, $B_\mathrm{GT}^{(n,p)}$ must be proportional to $\delta^2$, and from Eq. \eqref{eq:Ikedag}, $B_\mathrm{GT}^{(p,n)}$ cannot contain any terms linear in $\delta$.   We conclude that the small sum-rule contribution linear in $\delta$ in the energy-independent sum rule (first line of Table \ref{tab:sumrule}) reflects the fact that in a more realistic calculation, the neutron shell still remains mostly closed, with typically just a couple of neutrons excited into the $1g_{9/2}$ orbital. 

The results in Fig. \ref{fig:response_functions_fp}, shown for the three interactions, are reassuring.  In each calculation we see the same qualitative trends apparent in Fig. \ref{fig:response_functions_fpg}, with the tensor interaction leading to modest destructive interference at the lowest energies, $\lesssim 3$ MeV, and more substantial constructive interference in the region of $4$--$8$ MeV.  The results obtained from the two spaces differ systematically only at higher energies, near the GT resonance peak $\approx 12$--$14$ MeV, where the GT-complete SM calculations find modest destructive interference, while SM $2p_{3/2}1f_{5/2}2p_{1/2}1g_{9/2}$ interference is constructive.  At these energies, one expects configurations omitted from the $2p_{3/2}1f_{5/2}2p_{1/2}1g_{9/2}$ SM response calculation to play a significant role.

\begin{figure}
\centering
\includegraphics[scale=0.6]{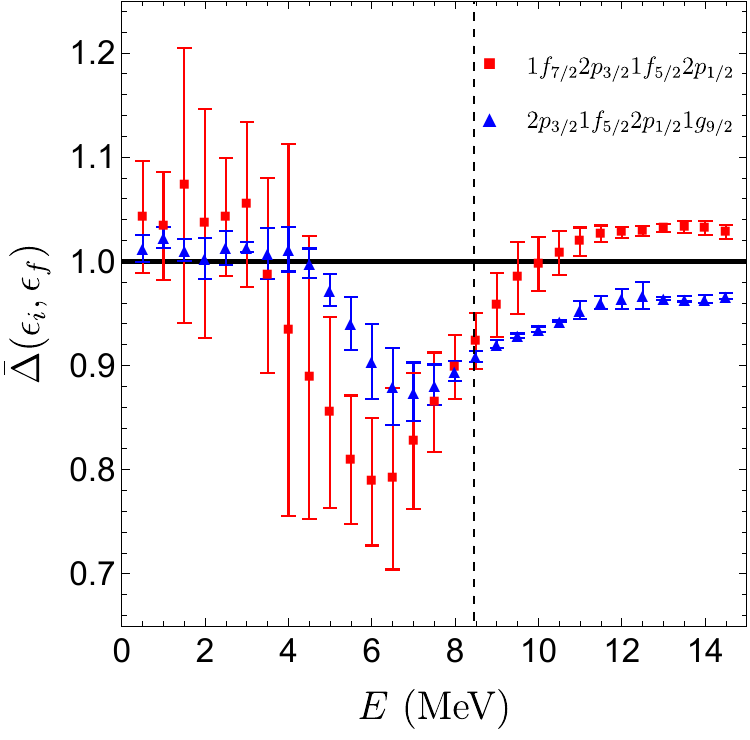}
\caption{The tensor correction $\Delta(\epsilon_i,\epsilon_f)$, Eq. \eqref{eq:ratio}.  A Lorentzian width $\Gamma=0.5$ MeV has been used, matching the bins $\epsilon_f-\epsilon_i = 0.5$ MeV used experimentally. Blue triangles (red squares) correspond to SM calculations performed in the $fpg$ ($fp$) model space. The vertical dashed line marks the energy $S_{n}(2^+)=8.46$ MeV, above which the branching ratio to the ground state of $^{71}$Ge is effectively zero. Error bars denote 1$\sigma$ uncertainties resulting from averaging the three SM calculations performed in each model space.
}
\label{fig:MT_SM}
\end{figure}
 
 Finally, we examine the ratio of the $\hat{O}_\mathrm{T^+}$ and $\hat{O}_\mathrm{GT^+}$ matrix elements, averaged over the energy bins of Table \ref{tab:bgt_krofcheck}, for the two SM spaces.  This is displayed in Fig. \ref{fig:MT_SM}.  Each SM result is assigned an uncertainty, computed from the $1\sigma$ variations among the three competing interactions we employ in each SM space.  This ratio is the underlying quantity we use to extract $B_\mathrm{GT^+}$ from $B_\mathrm{GT}^{(p,n)}$ via Eq. \eqref{eq:ratio}.  Below $\approx 8.5$ MeV --- the region relevant to solar neutrino studies ---  the two calculations are in reasonable agreement, with deviations generally smaller than the spreads induced by the choice of effective interactions used in each space.  Presumably the differences seen above 8.5 MeV reflect the impact of the $1g_{9/2}$ shell.
 
This crosscheck helps reassure us that the tensor distortions of the spectrum below 8.5 MeV are robust, and not an artifact
of the absence of spin-orbit partners in the $2p_{3/2}1f_{5/2}2p_{1/2}1g_{9/2}$-space Lanczos calculations.

\subsection{$B_\mathrm{GT^+}$ distribution}
Table \ref{tab:bgt_krofcheck} gives the extracted $B_\mathrm{GT^+}$ for nuclear excitation energies between
0.6 and 8.46 MeV as a binned distribution.  The bins have a typical width $\Delta E= E_f-E_i\approx 0.5$ MeV, 
and due to the very high density of nuclear states, represents the net contributions of many transitions.

Although a very good approximation, Bahcall simplified the sum over excited states by assigning the strength of
each bin to a ``doorway state'' placed at the midpoint of the bin; we follow experiment more faithfully by distributing the $B_\mathrm{GT^+}$ strength uniformly over each interval. Nuclear excitation energies become continuous, reflecting the high
density of states, with the $B^k_\mathrm{GT^+}$ for each bin $k$ defining a uniform density
\begin{equation}
    B^k_\mathrm{GT^+} = \int_{E^k_i}^{E^k_f} \left[\frac{ B^k_\mathrm{GT^+}}{\Delta E^k} \right] \, dE \equiv \int_{E^k_i}^{E^k_f} \rho^k_{B_\mathrm{GT^+}}  \, dE,
\end{equation}
with $E$ the nuclear excitation energy in $^{71}$Ge.  The sum and integration in Eq. (\ref{eq:sigmaor}) then
become 
\begin{equation}
  \langle \sigma \rangle \sim   \sum_f \int dE_\nu \, B_\mathrm{GT^+}\ldots \rightarrow \int_{Q^\mathrm{eff}}^{E_\nu^\mathrm{max}} dE_\nu \, \int_0^{E_\nu-Q^\mathrm{eff}} dE \,\rho_{B_\mathrm{GT^+}}\ldots,
  \label{eq:phasespace}
\end{equation}
where $\rho^k_{B_\mathrm{GT^+}}$ vanishes except when $E^k_i < E <E^k_f$ and
\[ \rho_{B_\mathrm{GT^+}} \equiv \sum_k \rho^k_{B_\mathrm{GT^+}} . \]
The sum extends over the 16 bins of Table \ref{tab:bgt_krofcheck}, restricting $E$ to values between 0.6 and 8.46 MeV.  In
Eq. (\ref{eq:phasespace}), $E_\nu^\mathrm{max}$ is the maximum neutrino energy of the solar source.  One adds to this result
the contributions of the first three discrete states of Table \ref{tab:bgt_krofcheck}, using Eq. (\ref{eq:sigmaor}) directly.

We take $B_\mathrm{GT}^{(p,n)}$ as given by experiment, in contrast to Bahcall who
scaled the experimental $B_\mathrm{GT}^{(p,n)}$ profile of \cite{PhysRevLett.55.1051} so that the ground-state $B_\mathrm{GT}^{(p,n)}$ value exactly matches that from EC.  There are several reasons we did not follow
Bahcall's procedure. First, the measured ground-state $B_\mathrm{GT}^{(p,n)}$ values, $0.089 \pm 0.007$ \cite{Krof87} and 
$0.085 \pm 0.015$ \cite{PhysRevLett.55.1051}, agree at $1 \sigma$ with the EC value extracted here, 0.0863. Second, the $(p,n)$ normalization used experimentally is unrelated to the ground state, based instead on the isospin analog state and the empirical relationship between
Fermi and GT strengths, determined from cases where both are known \cite{PhysRevC.25.1094}.  If this relationship were updated to take into account the current value
of $g_A$, the extracted $B_\mathrm{GT}^{(p,n)}$ values become $0.085 \pm 0.007$ and $0.082 \pm 0.015$, respectively, still in
agreement with the EC result.  Third, Bahcall's procedure equates EC and 
$(p,n)$ strengths and thus does not take into account the effect of the tensor amplitude.  Using the SM to estimate the tensor contribution in order to extract $B_\mathrm{GT^+}$ from the $(p,n)$ cross sections of \cite{Krof87,PhysRevLett.55.1051},
we find constructive interference and values of $0.079 \pm 0.006$ and $0.075 \pm 0.013$, again rougly consistent with the EC value at $1\sigma$. Furthermore, in our work we assign an overall 12\% normalization uncertainty to the measured
$B_\mathrm{GT}^{(p,n)}$, which exceeds any difference in the ground state EC and $(p,n)$
strengths, regardless of what corrections are included.

\section{Solar Neutrino Cross Section Results}
\label{sec:cs}
With the input neutrino spectra, cross section formulation, and needed $B_\mathrm{GT+}$ strengths now specified, we can compute
capture cross sections for the various solar neutrino sources. While the pp neutrino cross section is effectively determined by the EC rate, in other cases we must rely on transition strengths extracted
from the $B_\mathrm{GT}^{(p,n)}$ results of Table \ref{tab:bgt_krofcheck}. As noted previously, apart from the first two excited states, the needed information is available only as strength integrated over energy bins.  This is not a significant issue,
as the bins are narrow compared to the $^8$B and hep spectra. 

As discussed in \cite{PhysRevC.108.035502}, there are additional uncertainties affecting the overall normalization of the $(p,n)$ measurements that arise from the experimental treatment of efficiencies, beam normalization, neutron attenuation, background subtractions, etc. Based on past experimental evaluations of these effects, an estimate of the normalization uncertainty of $\pm$ 12\%(1$\sigma$)  was obtained in \cite{PhysRevC.108.035502}.  With the inclusion of this systematic normalization uncertainty, it was found that the overall uncertainty in the fitted value of $\delta=0.074\pm 0.008$ --- which employed data from a range of nuclei --- was also $\approx 12\%$. This is significant, as the weakest transitions providing the most purchase on $\delta$ would lead to larger variations in $\delta$, had the normalization uncertainty been underestimated.  Thus the experimental estimate of a 12\% normalization uncertainty appears reasonable from this statistical test.  We adopt the same normalization uncertainty here, treating it as a systematic uncertainty in that all of the bins grouping unresolved states are allowed to move up or down by this amount.  In general, this systematic uncertainty dominates for energetic solar neutrino sources that require integrations over many bins.

\subsection{\textnormal{pp} Neutrinos}
The driving reaction of the pp chain,  $p+p\rightarrow\, ^2\mathrm{H}+e^++\nu_e~$, 
produces a spectrum of neutrinos with endpoint energy  $E_{\nu,\mathrm{max}}^{\odot}(\mathrm{pp})=423.41\pm 0.03~\mathrm{keV}$, 
extending $\approx$ 3.2 keV beyond the laboratory value due to thermal effects described earlier.  The minimum neutrino energy for producing
$^{71}$Ge is set by the Q value,
$E_{\nu,\mathrm{min}}^{\odot}(\mathrm{pp})=232.5~\mathrm{keV}$.  

Using Eqs. \eqref{eq:sigmaor} and \eqref{eq:Beff} and including only the ground-state transition
\begin{equation}
    \begin{split}
\langle  \sigma \rangle_\mathrm{pp}^\mathrm{g.s.}  &= \frac{G_\mathrm{F}^2 \cos^2{\theta_\mathrm{C}} }{  \pi}    \int_{E_{\nu,\mathrm{min}}^{\odot}}^{E_{\nu,\mathrm{max}}^{\odot}}  dE_\nu~ P_\odot(E_\nu,W^\mathrm{th}_0)~  p_e E_e  \\
&\times {F}_{\beta^-}(Z_f,E_e)~g_A^2~B_{\mathrm{GT}^+}(\mathrm{g.s.}) \frac{[1+g_{v,b} ]_{(\nu,e)} }{[1+g_{v,b} ]_\mathrm{EC}} \frac{  [1+ \epsilon_q]_{(\nu,e)} }{   [1+ \epsilon_q]_\mathrm{EC} }, \nonumber \\
~
 \label{eq:pp}
  \end{split}
 \end{equation}
where the last two terms, treated as in \cite{PhysRevC.108.035502}, are the corrections for the differential effects of radiative corrections and weak magnetism on EC and pp neutrino capture.

As the na\"ive size of these corrections is $\approx$ 0.5\%, they were included in \cite{PhysRevC.108.035502} because the EC rate is known so precisely (now to $\lesssim$ 0.1\%). In fact, the two corrections were found to have opposite signs, leading to a net correction of just 0.2\%.  Consequently they play a minor role in the error budget for extracting $B^\mathrm{eff}_\mathrm{GT^+}$ from the EC rate. Nevertheless, for consistency with the earlier work, we have made the same corrections here, combining
Eq. \eqref{eq:Beff} with calculations of the radiative and weak-magnetism corrections for pp neutrino capture. We take the normalized pp neutrino spectrum from Bahcall \citep{Bahcall:1997eg}.  As the average energy for this spectrum is $\braket{E^\odot_\nu(\mathrm{pp})}=266.8~\mathrm{keV}$, only neutrinos in the upper half of the spectrum are above the threshold for capture.
Performing the integral in Eq. \eqref{eq:pp} we find
\begin{equation}
\langle \sigma \rangle_\mathrm{pp}^\mathrm{g.s.} =\left(1.158\pm 0.009\right)\times 10^{-45}~\mathrm{cm}^2 
\label{eq:ppfinal}
\end{equation}
There is an additional contribution from the $5/2^-$ first excited state of $^{71}$Ge, but only neutrinos within 16 keV of the endpoint are sufficiently energetic to drive this transition.  Extracting the
$B_\mathrm{GT^+}$ value from $B_\mathrm{GT}^{(p,n)}$ (see Table \ref{tab:bgt_krofcheck}), we find that the contribution is negligible.  Thus $\langle \sigma \rangle_\mathrm{pp} \approx \langle \sigma \rangle_\mathrm{pp}^\mathrm{g.s.}$
to the precision given in Eq. \eqref{eq:ppfinal}.

For other transitions, uncertainties are larger, dominated by the extraction of $B_\mathrm{GT^+}$ from $(p,n)$ results; radiative corrections are neglected in these cases. We do include weak magnetism corrections in the $^8$B and hep calculations, where momentum transfers are much larger.  The procedure for doing so 
is described below.

\subsection{\texorpdfstring{$^7\mathrm{Be}$}{7Be} Neutrinos}
The $^7$Be neutrinos are the second most important solar source.  The EC reaction 
\begin{equation}
^7\mathrm{Be}+e^-\rightarrow\,^7\mathrm{Li}+ \nu_e
\end{equation}
proceeds dominantly to the $3/2^-$ ground state of the daughter nucleus with $Q_\mathrm{EC}=861.8$ keV and to the $1/2^-$ excited state at 477.612 keV with branching ratio 10.44\%, producing two distinct line sources of neutrinos. As discussed in Sec. \ref{sec:spectra}, thermal motion within the sun broadens each line asymmetrically,
generating a width at half maximum of  $\approx 1.7$ keV \cite{PhysRevD.49.3923},  while shifting the average line energies by
\begin{equation}
\begin{split}
\braket{E_{\nu}^\odot(^7\mathrm{Be})}&=861.8~\mathrm{keV}+1.28~\mathrm{keV}=863.1~\mathrm{keV},\\
\braket{E_{\nu}^\odot(^7\mathrm{Be})}&=384.3~\mathrm{keV}+1.24~\mathrm{keV}=385.5~\mathrm{keV}.
\end{split}
\end{equation}

The 385 keV line can only excite the transition to the ground state of $^{71}$Ge, whereas the 861 keV neutrinos can excite the $5/2^-$ and $3/2^-$ states at 175 and 500 keV, respectively.  Both of these excited states were resolved in the forward-angle $(p,n)$ measurements of Krofcheck et al. \cite{PhysRevLett.55.1051,Krof87}.  The calculation of the $^7$Be cross section thus follows very closely that performed for the $^{37}$Ar and $^{51}$Cr 
neutrino sources in \cite{PhysRevC.108.035502}, where the same two excited states contribute.  In the analysis described there,  the  $B_\mathrm{GT^+}$ values
were extracted from the $(p,n)$ results, accounting for the tensor operator contributions and for SM uncertainties in evaluating these contributions, 
propagating errors by Monte Carlo yielding the results in Eq. \eqref{eq:resolved_BGTs}. We use these results, referring the reader to the original source for the detail. We find for the cross section
\begin{equation}
\braket{\sigma}_{^7\mathrm{Be}}=7.05^{+0.35}_{-0.09}\times 10^{-45}~\mathrm{cm}^2.
\end{equation}
The calculation treats the thermally broadened $^7$Be neutrino lines by integrating over the numerical profiles given by Bahcall \cite{PhysRevD.49.3923}, 
though omission of the line broadening does not alter the result above.

\subsection{\textnormal{pep} Neutrinos}
Neutrinos produced via the pep reaction
\begin{equation}
p+p+e^-\rightarrow\,^2\mathrm{H}+\nu_e~,
\end{equation}
are well approximated as a line source, which after accounting for thermal effects has an energy 
\begin{equation}
E_{\nu}^{\odot}(\mathrm{pep})=1.447~\mathrm{MeV}.
\end{equation}
Consequently, pep neutrinos have sufficient energy to reach $^{71}$Ge states up to 1.214 MeV. Based on the level assignments of \cite{SINGH20231}, there are nine candidate GT transitions to states above 500 keV, with energies ranging from 0.706 to 1.212 MeV.  Transitions to these states were not resolved in the
$(p,n)$ studies of \cite{PhysRevLett.55.1051,Krof87}, but instead are grouped into the bins of Table \ref{tab:bgt_krofcheck}. We use the BGT strengths from the first two bins, which are centered at 0.8 and 1.25 MeV, treating the strengths within each bin as uniformly distributed.  That is, only a fraction of the $B_\mathrm{GT^+}$ strength within the $1.0$--$1.5$ MeV bin is included.

We find
\begin{equation}
\braket{\sigma}_\mathrm{pep}=20.4^{+1.0}_{-0.5}\times 10^{-45}~\mathrm{cm}^2.
\end{equation}
The unresolved states contribute ($12\pm 2$)\% of this total.

\subsection{\textnormal{CNO} Neutrinos}
The CNO $\beta$-decay reactions 
\begin{equation}
\begin{split}
^{13}\mathrm{N}&\rightarrow\,^{13}\mathrm{C}+e^++\nu_e,\\
^{15}\mathrm{O}&\rightarrow\,^{15}\mathrm{N}+e^++\nu_e, \\
^{17}\mathrm{F}&\rightarrow\,^{17}\mathrm{O}+e^++\nu_e,
\end{split}
\end{equation}
produce allowed neutrino spectra with end-point energies
\begin{equation}
\begin{split}
E_{\nu,\mathrm{max}}^{\odot}(^{13}\mathrm{N})&=1.1982\pm 0.0003~\mathrm{MeV},\\
E_{\nu,\mathrm{max}}^{\odot}(^{15}\mathrm{O})&=1.7317\pm 0.0005~\mathrm{MeV},\\
E_{\nu,\mathrm{max}}^{\odot}(^{17}\mathrm{F})&=1.7364\pm 0.0003~\mathrm{MeV},
\end{split}
\end{equation}
and thus can excite states in $^{71}$Ge below 0.796, 1.499, and 1.504 MeV, respectively.  Consequently, unresolved states grouped within the first
bin of Table \ref{tab:bgt_krofcheck} contribute to the capture of all CNO sources, while those in the second bin contribute to the capture of $^{15}$O and
$^{17}$F neutrinos. 

We adopt the allowed CNO neutrino spectra of Bahcall \cite{Bahcall:1997eg} and perform the phase-space integrations to find
\begin{equation}
\begin{split}
\braket{\sigma}_{^{13}\mathrm{N}}&=5.94^{+0.28}_{-0.07}\times 10^{-45}~\mathrm{cm}^2, \\
\braket{\sigma}_{^{15}\mathrm{O}}&=11.2^{+0.5}_{-0.2}\times 10^{-45}~\mathrm{cm}^2,\\
\braket{\sigma}_{^{17}\mathrm{F}}&=11.2^{+0.5}_{-0.2}\times 10^{-45}~\mathrm{cm}^2.
\end{split}
\end{equation}
The ground-state transition, with its precisely known $B_\mathrm{GT^+}$ strength, accounts for $91$, $84$, and $84$\%, respectively, of the total cross section values.   The influence of the unresolved excited states is somewhat diminished by the $\beta$-decay phase space, with these states accounting for $(1.8\pm 0.3)$, $(8\pm 1)$, and $(8\pm 1)$\% of the totals, respectively.  Thus in each case the excited-state contribution is roughly split between the unresolved states and the resolved 175 and 500 keV states, with the former dominating the resulting uncertainty bands.

\subsection{\texorpdfstring{$^8\mathrm{B}$}{8B} Neutrinos}
The energetic solar neutrinos produced in the $\beta$ decay of $^8$B, 
\begin{equation}
^8\mathrm{B}\rightarrow\,^8\mathrm{Be}^*+e^++\nu_e,
\end{equation}
are of special importance because of their sensitivity to the solar core temperature and because they are the
branch observed in high-threshold detectors like SuperKamiokande.  The spectrum extends to approximately 17 MeV and
deviates from an allowed shape because the final state in $^8$Be is a broad resonance.  The results of several recent
measurements that have helped to better define the spectrum are summarized in SFIII \cite{Acharya:2024lke}.
We follow the recommendations in this review to adopt the experimentally derived spectrum of Longfellow et al. \cite{PhysRevC.107.L032801}.
The procedures followed in \cite{PhysRevC.107.L032801} take into account weak magnetism (the interference term between
vector and axial-vector three-currents) and radiative corrections.

Because of the higher energies of these neutrinos, we also include in the cross sections the
weak magnetism contribution.  This yields
\begin{equation}
\begin{split}
&\langle  \sigma \rangle_{^8\mathrm{B}}^\mathrm{g.s.}  = \frac{G_\mathrm{F}^2 \cos^2{\theta_\mathrm{C}} }{  \pi}  \\
&\times \sum_{f \in \mathrm{bound}}\int  dE_\nu~ P^\mathrm{Long}_\odot(E_\nu)~ p_e E_e~{F}_{\beta^-}(Z_f,E_e)  ~ g_A^2~B^\mathrm{WM}_\mathrm{GT^+},  \\
\end{split}
 \label{eq:8B}
 \end{equation}
where the sum extends over all bound and quasi-bound states in $^{71}$Ge, as we described previously, and
\begin{equation}
\begin{split}
&B^\mathrm{WM}_\mathrm{GT^+} \\
&\equiv B_\mathrm{GT^+} \left[ 1 + \frac{2 \bar{E} }{ 3 m_N g_A} \left(  \mu_{T=1} + \frac{ \langle j_f \alpha_f || \sum_{i=1}^A \boldsymbol{\ell}(i) \tau_+(i) || j_i \alpha_i \rangle }{  \langle j_f \alpha_f || \sum_{i=1}^A \boldsymbol{\sigma}(i) \tau_+(i) || j_i \alpha_i \rangle} \right)  \right],
\end{split}
\end{equation}
with $m_N$ being the nucleon mass and $\bar{E}\equiv E_e +E_\nu-m_e^2/E_e$.  
 
In our convention all matrix elements are real. Because the isovector magnetic moment $\mu_{T=1} = 4.706 \gg 1$, 
one expects the first term in the correction to dominate over the second.  Qualitative arguments have been made that the second term can be 
absorbed into the first by the replacement
\[ \frac{ \langle j_f \alpha_f || \sum_{i=1}^A \boldsymbol{\ell}(i) \tau_+(i) || j_i \alpha_i \rangle }{ \langle j_f \alpha_f || \sum_{i=1}^A \boldsymbol{\sigma}(i) \tau_+(i) || j_i \alpha_i \rangle} \rightarrow - \frac{1}{ 2} .\]
While this assumption is motivated by studies of transitions between spin-orbit partners, its validity has been tested recently in shell-model calculations, where on average it has been found to hold, though with significant deviations correlated with cases where the spin matrix element is weak \cite{PhysRevC.95.064313}.  Such cases, though, contribute
little to the semi-inclusive sum of interest, because the correction factor multiplies $B_\mathrm{GT+}$.   We took the further precaution of testing this simplification
by computing the associated energy-independent sum rules for $^{71}$Ga.  For the SM space $2p_{3/2}1f_{5/2}2p_{1/2}1g_{9/2}$, we obtained a sum-rule average of -0.53.  
Thus we make the simplification
\begin{equation}
B^\mathrm{WM}_\mathrm{GT^+} \approx B_\mathrm{GT^+} \left[ 1 + \frac{2 \bar{E} }{ 3 m_N g_A} \left(  \mu_{T=1} - \frac{1 }{2} \right)  \right].
\label{eq:WM}
\end{equation}

Folding the corrected cross section with the Longfellow spectrum and summing over the $B_\mathrm{GT^+}$ strengths of Table \ref{tab:bgt_krofcheck} we obtain
\begin{equation}
\braket{\sigma}_{^8\mathrm{B}}=2.57^{+0.30}_{-0.25}\times 10^{-42}~\mathrm{cm}^2.
\end{equation}
The unresolved excited states dominate this result, accounting for $(87\pm 1) \%$ of the total. Likewise, the $\pm 12\% $ uncertainty in the normalization of the $(p,n)$ data is the dominant source of uncertainty.

\subsection{\textnormal{hep} Neutrinos}
A rare branch of the pp chain produces the high-energy hep neutrinos
\begin{equation}
^3\mathrm{He}+p\rightarrow\,^4\mathrm{He}+e^++\nu_e~,
\end{equation}
with a maximum energy of 18.778  MeV.  As discussed in SFIII \cite{Acharya:2024lke}, there is no adequate
discussion of the shape of this spectrum in the literature.  While nominally an allowed transition, the GT operator 
cannot connect the dominant $s$-wave components of the initial $p+^3$He and final $^4$He states.  Consequently,
weak magnetism and other one-body corrections to the vector current, two-body vector and axial-vector currents, 
and $p$-waves in the $p+^3$He wave function, normally unimportant under solar conditions, are significantly enhanced.  In addition,
cancellations arise between the one- and two-body contributions.  While all of these effects have been
addressed in sophisticated few-body calculations of the $S$ factor \cite{PhysRevC.63.015801,PhysRevC.67.055206},
the publications did not provide the resulting neutrino spectrum. In the absence of a modern spectrum,  we continue to use the allowed spectrum of Bahcall \cite{Bahcall:1997eg}.

Folding this spectrum with the $B_\mathrm{GT^+}$ strengths of Table \ref{tab:bgt_krofcheck} while using Eqs. \eqref{eq:8B}, \eqref{eq:WM} yields
\begin{equation}
\braket{\sigma}_\mathrm{hep}=7.84^{+0.9}_{-0.9}\times 10^{-42}~\mathrm{cm}^2.
\end{equation}
The unresolved excited states also dominate this result, accounting for $(92\pm 1)\%$ of the total.

\begin{table*}
\centering
{\renewcommand{\arraystretch}{1.4}
\begin{tabular}{cccccc}
\hline
\hline
~~~~ & ~$\langle \sigma \rangle_\mathrm{HR}$ ~ & ~$\langle \sigma \rangle_\mathrm{Bahcall97}$ ~ & ~~$\langle \sigma \phi \rangle_\mathrm{HR,GS98}$~~ & ~~$\langle \sigma \phi \rangle_\mathrm{HR,AAG21}$~~ & ~~$\langle \sigma \phi \rangle_\mathrm{HR,MB22p}$~~\\
 ~~$\nu$ Source~~ &($10^{-45}$ cm$^2$)  & ($10^{-45}$ cm$^2$) & (SNU) & (SNU) & (SNU) \\
\hline
pp & $1.158^{+0.009}_{-0.009}$  & $1.172^{+0.027}_{-0.027}$ & $69.0^{+0.7}_{-0.7}$ & $69.5^{+0.7}_{-0.7}$ & $68.9^{+0.7}_{-0.7}$\\
pep & $20.4^{+1.0}_{-0.5}$ & $20.4^{+3.5}_{-1.4}$ & $2.92^{+0.15}_{-0.08}$ & $2.96^{+0.15}_{-0.08}$ & $2.90^{+0.15}_{-0.08}$\\
$^7$Be & $7.05^{+0.35}_{-0.09}$ & $7.17^{+0.50}_{-0.22}$ & $34.2^{+3.0}_{-2.6}$ & $31.9^{+2.8}_{-2.4}$ & $34.4^{+3.3}_{-2.8}	$ \\
$^8$B &  $2.57^{+0.30}_{-0.25}\times 10^{3}$  & $2.40^{+0.77}_{-0.36}\times 10^{3}$ & $12.9^{+2.3}_{-2.3}$ & $11.1^{+1.9}_{-1.9}$ & $13.0^{+2.5}_{-2.5}$ \\
hep &  $7.84^{+0.9}_{-0.9}\times 10^3$  & $7.14^{+2.3}_{-1.1}\times 10^3$ & $0.062^{+0.021}_{-0.021}$ & $0.064^{+0.021}_{-0.021} $ & $0.062^{+0.020}_{-0.020}$\\
$^{13}$N & $5.94^{+0.28}_{-0.07}$  & $6.04^{+0.36}_{-0.18}$ & $1.66^{+0.28}_{-0.27}$ & $1.32^{+0.18}_{-0.17}$ & $1.84^{+0.29}_{-0.28}$\\
$^{15}$O & $11.2^{+0.5}_{-0.2}$  & $11.4^{+1.4}_{-0.6}$ & $2.32^{+0.43}_{-0.42}$ & $1.77^{+0.29}_{-0.28}$ & $2.58^{+0.48}_{-0.47}$ \\
$^{17}$F & $11.2^{+0.5}_{-0.2}$ & $11.4^{+1.4}_{-0.6}$ & $0.060^{+0.012}_{-0.012}$ & $0.038^{+0.006}_{-0.006}$ & $0.053^{+0.009}_{-0.009} $ \\
Total & & & $123.2^{+6.9}_{-6.3}$   & $118.6^{+6.1}_{-5.5}$  & $123.8^{+7.4}_{-6.8}$  \\
\hline
\hline
\end{tabular}
}
\caption{Comparison of the $^{71}$Ga neutrino capture cross sections determined here vs. those of Bahcall \cite{Bahcall:1997eg}.  Capture rates in solar neutrino units (1 SNU $\equiv  10^{-36}$ captures/$^{71}$Ga atom/s) are obtained
by folding the updated cross sections 
with the SSM fluxes and their uncertainties, as given in Table \ref{tab:fluxes}.   The uncertainties on the total rate are very conservatively
obtained by adding the uncertainties for each separate capture rate, thereby treating these rates as maximally correlated.}
\label{tab:solar_rates}
\end{table*}

\subsection{Cross Section Summary}
The results of this section are summarized in Table \ref{tab:solar_rates} and compared to 
\cite{Bahcall:1997eg}. Bahcall's 1997 cross sections remained in use throughout the lifetimes of the SAGE and GALLEX/GNO experiments, employed in the analyses that constrained the solar neutrino flux.  Consequently, these cross sections are embedded in modern global neutrino analyses that use the constraints from these archival experiments. Table \ref{tab:solar_rates} also includes the $^{71}$Ga capture rate $\langle \sigma \phi \rangle$ for the fluxes of the three SSMs of Table \ref{tab:fluxes}.

Our analysis narrows uncertainties on the various solar neutrino cross sections significantly.  However, with the exception of the pp
cross section, Bahcall's central values fall within our $1\sigma$ error bands. (The pp exception reflects the precise value of $g_A$ now available.) The largest
differences between the present calculation and \cite{Bahcall:1997eg} are found for the high-energy $^8$B and hep neutrinos, where our cross section central values are $\approx 7\%$ and $10\%$ larger than those of Bahcall, respectively. Key differences in the calculations include:
\begin{enumerate}
    \item The contribution of continuum bins below 8.46 MeV, where $\Gamma_\gamma/\Gamma$ is significant.  This increases our $^8$B cross section by $\approx 4\%$ (before accounting for the tensor-operator correction).
    \item The tensor operator. Below 4 MeV the interference of the GT and tensor operators is destructive, so the correction increases the extracted $B_\mathrm{GT^+}$
    strength, and thus the $^8$B cross section, by $0.3\%$.  Above 4 MeV the interference is constructive, so the correction decreases the extracted $B_\mathrm{GT^+}$ strength. The tensor effects are large, but the phase space for these states is less favorable, leading to a decrease in the $^8$B cross
    section of $-2.5\%$.  The sum is a net decrease of $-2.2\%$.
     \item The weak magnetism correction, which increases our $^8$B cross section estimate by $\approx 2\%$.  Bahcall discusses
     this correction, though his numerical procedures for $^{71}$Ga (including treatment of the orbital angular momentum contribution) are unclear:
     he references for details a calculation of $^8$B $\beta$ decay \cite{PhysRevC.33.2121}. Thus we are unable to assess whether
     the two treatments of weak magnetism are in quantitative agreement.

    \item The use of the $^8$B neutrino spectrum determined by Longfellow et al. \cite{PhysRevC.107.L032801} increases the capture rate by $\approx 2.4\%$.
     
     \item Bahcall's rescaling of the experimental $B_\mathrm{GT}^{(p,n)}$ values.  The normalizing ground-state EC $B_\mathrm{GT^+}$ is not given explicitly in \cite{Bahcall:1997eg}, but we estimate his procedures would have reduced the transition strengths by $\approx$ 2\%, thus diminishing his cross sections relative to ours by the same amount.
\end{enumerate}

\section{Summary}
The cross sections derived here and presented in Table \ref{tab:solar_rates} are only somewhat different than those recommended by Bahcall \cite{Bahcall:1997eg}.
The central-value changes range from $\approx -2\%$ for $^7$Be and CNO neutrinos to $\approx +10\%$ for hep neutrinos. 
As noted above, in general our central values are in good agreement with those of Bahcall, which typically fall within the 
$1\sigma$ uncertainties we have derived --- though in some cases this comes about because of cancellations among the
various corrections we have included.  Our uncertainties, which are significantly smaller than those of \cite{Bahcall:1997eg},
are derived by quantitatively evaluating errors on parameter inputs, such as the $(p,n)$ cross section uncertainties, tensor
operator uncertainties, and any input needed from nuclear microphysics, then propagating these uncertainties by Monte Carlo
through the calculations.  In contrast, the uncertainty analysis of \cite{Bahcall:1997eg} was significantly less quantitative,
as described below.

For the high-energy hep and $^8$B neutrinos, the largest change in the total cross section comes from including contributions from two energy bins in the continuum where $\Gamma_\gamma/\Gamma$ is not negligible.  Bahcall assumed that continuum states would not
decay electromagnetically and thus would not yield countable $^{71}$Ge. This increase in the cross section is partially canceled by the tensor correction, which interferes constructively with the GT amplitude for excitations in the range of $4$--$8$ MeV, thus modestly reducing the extracted $B_\mathrm{GT+}$ strengths.

The uncertainties we obtain are typically reduced by a factor of $\approx 2$--$3$, compared to those of \cite{Bahcall:1997eg}. In the case of the pp neutrinos, the uncertainty reduction is the net result of our use of modern weak couplings with reduced errors, explicit calculations of small corrections due to weak magnetism and radiative corrections, and improvements in the measured EC rate and Q value.

For higher-energy neutrinos that induce transitions to excited states of $^{71}$Ge, the improved uncertainties reflect our more sophisticated extraction of GT strengths from charge-exchange data. In his work, Bahcall equated $B_\mathrm{GT^+}$ to $B_\mathrm{GT}^{(p,n)}$, but having no quantitative way to address uncertainties in this relationship, arbitrarily designated a $3\sigma$ uncertainty in extracted $B_\mathrm{GT^+}$ strengths by either doubling or halving the experimental values for $B_\mathrm{GT}^{(p,n)}$. Bahcall's $1\sigma$ cross-section uncertainties, reproduced here in Table \ref{tab:solar_rates}, were taken to be one-third of the resulting $3 \sigma$ uncertainties.  Effectively, his uncertainty sweeps questions about the relationship between $B_\mathrm{GT^+}$ and $B_\mathrm{GT}^{(p,n)}$ into a correction analogous to our normalization uncertainty, taking that to be 33\% at $1\sigma$.

However, as discussed in more detail here and in \cite{PhysRevC.108.035502}, a better understanding of the relationship between GT transitions and forward-angle $(p,n)$ amplitudes has been achieved by testing the latter in transitions where the $B_\mathrm{GT+}$ strengths are known.  The inclusion of the subdominant tensor operator in the $(p,n)$ analysis systematically accounts for the differences between the two probes.  Further, analyses of experimental systematics suggest a $1\sigma$ normalization uncertainty in $B_\mathrm{GT}^{(p,n)}$ of about 12\%, with this estimate supported by the correlation between the tensor-corrected $B_\mathrm{GT}^{(p,n)}$ and $B_\mathrm{GT^+}$, now that this correlation extends to weak transitions where $\delta$ is important, if not dominant.  As noted in \cite{PhysRevC.108.035502}, 12\% is consistent with many past estimates of the normalization uncertainty, which typically have ranged between 10\% and 20\%.  Our more nuanced understanding of the relationship of $B_\mathrm{GT}^{(p,n)}$ to $B_\mathrm{GT^+}$ has thus given us much more confidence in our assessment of the true uncertainties inherent in the use of forward-angle $(p,n)$ cross sections to map out GT strengths.

We believe the current effort is the first to explore the effects of the tensor correction in the effective operator for forward-angle
$(p,n)$ scattering on extractions of the inclusive $B_\mathrm{GT}$ response.  In retrospect, it is regrettable that such investigations were not undertaken during the period when major experimental programs on forward-angle $(p,n)$ scattering, like that at IUCF \cite{PhysRevLett.44.1755}, were underway.  While we have
focused on the task at hand --- better defining the $^{71}$Ga solar neutrino cross section --- the phenomenology uncovered
suggests that the community did not fully recognize all of the opportunities to connect $(p,n)$ studies with shell structure.   The generalized Ikeda sum rule 
derived here is quadratic in the tensor coupling strength $\delta$ and thus little changed.  Effects linear in $\delta$ in the
energy-independent sum rule cancel precisely in the difference between $B^{(p,n)}_\mathrm{GT}$ and $B^{(n,p)}_\mathrm{GT}$.  Consequently,
in systems like $^{71}$Ga with a large neutron excess and largely blocked $B^{(n,p)}_\mathrm{GT}$, this implies that the linear contribution
to the energy-independent sum rule for $B^{(p,n)}_\mathrm{GT}$ must be very small.   We found precisely this behavior
in our sum rule evaluations.  But this cancellation is not what one expects in single-particle transitions, where na\"ively $\langle \sigma \rangle^2  \approx \langle \sigma \rangle  \langle \ell \rangle$.   That is, one expects deviations
between $B_\mathrm{GT^+}$ and $B_\mathrm{GT}^{(p,n)}$, and between $B_\mathrm{GT^-}$ and $B_\mathrm{GT}^{(n,p)}$, that are linear in $\delta$,
with signs fluctuating.  Thus the tensor correction leads to a distortion of the $B_\mathrm{GT}^{(p,n)}$ profile relative to $B_\mathrm{GT^+}$,
but not a significant increase in the overall integrated strength.  The distortions, including their varying signs, are determined 
by the underlying single-particle amplitudes (as in Table \ref{tab:M_ratios}) and by where the strength associated with those amplitudes concentrates
in the spectrum.  We see this behavior manifested in the truncated SM results of Table \ref{tab:sumrule}. The knowledge that the leading effect of the tensor correction for nuclei with largely unblocked valence neutrons is a spectral distortion, rather than a change in the energy-unweighted sum rule, opens up a role for nuclear theory in future forward-angle $(p,n)$ analyses.

\section*{Acknowledgements}
WH acknowledges support by the US Department of Energy under grants DE-SC0004658, DE-SC0023663, and DE-AC02-05CH11231, the National Science Foundation under cooperative agreement 2020275, and the Heising-Simons Foundation under award 00F1C7. ER is supported by the National Science Foundation under cooperative agreement 2020275 and by the U.S. Department of Energy through the Los Alamos National Laboratory. Los Alamos National Laboratory is operated by Triad National Security, LLC, for the National Nuclear Security Administration of U.S. Department of Energy (Contract No. 89233218CNA000001).

\bibliographystyle{elsarticle-num}
\bibliography{CS2}

\begin{thebibliography}{10}
\expandafter\ifx\csname url\endcsname\relax
  \def\url#1{\texttt{#1}}\fi
\expandafter\ifx\csname urlprefix\endcsname\relax\def\urlprefix{URL }\fi
\expandafter\ifx\csname href\endcsname\relax
  \def\href#1#2{#2} \def\path#1{#1}\fi

\bibitem{ANSELMANN1992376}
P.~Anselmann, W.~Hampel, G.~Heusser, J.~Kiko, T.~Kirsten, E.~Pernicka, R.~Plaga, U.~R{\"o}nn, M.~Sann, C.~Schlosser, R.~Wink, M.~W{\'o}jcik, R.~{von Ammon}, K.~Ebert, T.~Fritsch, K.~Hellriegel, E.~Henrich, L.~Stieglitz, F.~Weyrich, M.~Balata, E.~Bellotti, N.~Ferrari, H.~Lalla, T.~Stolarczyk, C.~Cattadori, O.~Cremonesi, E.~Fiorini, S.~Pezzoni, L.~Zanotti, F.~{von Feilitzsch}, R.~M{\"o}{\ss}bauer, U.~Schanda, G.~Berthomieu, E.~Schatzman, I.~Carmi, I.~Dostrovsky, C.~Bacci, P.~Belli, R.~Bernabei, S.~d'Angelo, L.~Paoluzi, S.~Charbit, M.~Cribier, G.~Dupont, L.~Gosset, J.~Rich, M.~Spiro, C.~Tao, D.~Vignaud, R.~Hahn, F.~Hartmann, J.~Rowley, R.~Stoenner, J.~Weneser, \href{https://www.sciencedirect.com/science/article/pii/037026939291521A}{Solar neutrinos observed by gallex at gran sasso}, Physics Letters B 285~(4) (1992) 376--389.
\newblock \href {https://doi.org/https://doi.org/10.1016/0370-2693(92)91521-A} {\path{doi:https://doi.org/10.1016/0370-2693(92)91521-A}}.
\newline\urlprefix\url{https://www.sciencedirect.com/science/article/pii/037026939291521A}

\bibitem{GALLEX:1998kcz}
W.~Hampel, et~al., {GALLEX solar neutrino observations: Results for GALLEX IV}, Phys. Lett. B 447 (1999) 127--133.
\newblock \href {https://doi.org/10.1016/S0370-2693(98)01579-2} {\path{doi:10.1016/S0370-2693(98)01579-2}}.

\bibitem{Kaether:2010ag}
F.~Kaether, W.~Hampel, G.~Heusser, J.~Kiko, T.~Kirsten, {Reanalysis of the GALLEX solar neutrino flux and source experiments}, Phys. Lett. B 685 (2010) 47--54.
\newblock \href {http://arxiv.org/abs/1001.2731} {\path{arXiv:1001.2731}}, \href {https://doi.org/10.1016/j.physletb.2010.01.030} {\path{doi:10.1016/j.physletb.2010.01.030}}.

\bibitem{GNO:2005bds}
M.~Altmann, et~al., {Complete results for five years of GNO solar neutrino observations}, Phys. Lett. B 616 (2005) 174--190.
\newblock \href {http://arxiv.org/abs/hep-ex/0504037} {\path{arXiv:hep-ex/0504037}}, \href {https://doi.org/10.1016/j.physletb.2005.04.068} {\path{doi:10.1016/j.physletb.2005.04.068}}.

\bibitem{SAGE:2009eeu}
J.~N. Abdurashitov, et~al., {Measurement of the solar neutrino capture rate with gallium metal. III: Results for the 2002--2007 data-taking period}, Phys. Rev. C 80 (2009) 015807.
\newblock \href {http://arxiv.org/abs/0901.2200} {\path{arXiv:0901.2200}}, \href {https://doi.org/10.1103/PhysRevC.80.015807} {\path{doi:10.1103/PhysRevC.80.015807}}.

\bibitem{Gavrin:2019sok}
V.~N. Gavrin, \href{https://www.worldscientific.com/doi/abs/10.1142/9789811204296_0002}{The history, present and future of SAGE (Soviet-American Gallium Experiment)}, World Scientific, 2019, pp. 29--46.
\newblock \href {http://arxiv.org/abs/https://www.worldscientific.com/doi/pdf/10.1142/9789811204296_0002} {\path{arXiv:https://www.worldscientific.com/doi/pdf/10.1142/9789811204296_0002}}, \href {https://doi.org/10.1142/9789811204296_0002} {\path{doi:10.1142/9789811204296_0002}}.
\newline\urlprefix\url{https://www.worldscientific.com/doi/abs/10.1142/9789811204296_0002}

\bibitem{PhysRevLett.20.1205}
R.~Davis, D.~S. Harmer, K.~C. Hoffman, \href{https://link.aps.org/doi/10.1103/PhysRevLett.20.1205}{Search for neutrinos from the sun}, Phys. Rev. Lett. 20 (1968) 1205--1209.
\newblock \href {https://doi.org/10.1103/PhysRevLett.20.1205} {\path{doi:10.1103/PhysRevLett.20.1205}}.
\newline\urlprefix\url{https://link.aps.org/doi/10.1103/PhysRevLett.20.1205}

\bibitem{Kamiokande-II:1989hkh}
K.~S. Hirata, et~al., {Observation of B-8 Solar Neutrinos in the Kamiokande-II Detector}, Phys. Rev. Lett. 63 (1989) 16.
\newblock \href {https://doi.org/10.1103/PhysRevLett.63.16} {\path{doi:10.1103/PhysRevLett.63.16}}.

\bibitem{PhysRevD.49.3622}
N.~Hata, S.~Bludman, P.~Langacker, \href{https://link.aps.org/doi/10.1103/PhysRevD.49.3622}{Astrophysical solutions are incompatible with the solar neutrino data}, Phys. Rev. D 49 (1994) 3622--3625.
\newblock \href {https://doi.org/10.1103/PhysRevD.49.3622} {\path{doi:10.1103/PhysRevD.49.3622}}.
\newline\urlprefix\url{https://link.aps.org/doi/10.1103/PhysRevD.49.3622}

\bibitem{PhysRevD.50.4749}
V.~Castellani, S.~Degl'Innocenti, G.~Fiorentini, M.~Lissia, B.~Ricci, \href{https://link.aps.org/doi/10.1103/PhysRevD.50.4749}{Neutrinos from the sun: Experimental results confronted with solar models}, Phys. Rev. D 50 (1994) 4749--4761.
\newblock \href {https://doi.org/10.1103/PhysRevD.50.4749} {\path{doi:10.1103/PhysRevD.50.4749}}.
\newline\urlprefix\url{https://link.aps.org/doi/10.1103/PhysRevD.50.4749}

\bibitem{1995ARA&A..33..459H}
W.~C. Haxton, {The Solar Neutrino Problem}, Annu. Rev. A \& A 33 (1995) 459--504.
\newblock \href {http://arxiv.org/abs/hep-ph/9503430} {\path{arXiv:hep-ph/9503430}}, \href {https://doi.org/10.1146/annurev.aa.33.090195.002331} {\path{doi:10.1146/annurev.aa.33.090195.002331}}.

\bibitem{Esteban:2020cvm}
I.~Esteban, M.~Gonzalez-Garcia, M.~Maltoni, et~al., The fate of hints: updated global analysis of three-flavor neutrino oscillations, J. High Energ. Phys. 09 (2020) 178.

\bibitem{deSalas:2020pgw}
P.~F. de~Salas, D.~V. Forero, S.~Gariazzo, et~al., 2020 global reassessment of the neutrino oscillation picture, JHEP 02 (2021) 071.

\bibitem{PhysRevD.104.083031}
F.~Capozzi, E.~Di~Valentino, E.~Lisi, A.~Marrone, A.~Melchiorri, A.~Palazzo, Unfinished fabric of the three neutrino paradigm, Phys. Rev. D 104 (2021) 083031.

\bibitem{zavatarelli20161753}
S.~Zavatarelli, G.~Bellini, J.~Benziger, D.~Bick, G.~Bonfini, D.~Bravo, B.~Caccianiga, L.~Cadonati, F.~Calaprice, A.~Caminata, P.~Cavalcante, A.~Chavarria, A.~Chepurnov, D.~D'Angelo, S.~Davini, A.~Derbin, A.~Empl, A.~Etenko, K.~Fomenko, D.~Franco, F.~Gabriele, C.~Galbiati, S.~Gazzana, C.~Ghiano, M.~Giammarchi, M.~GÃ¶ger-Neff, A.~Goretti, M.~Gromov, C.~Hagner, E.~Hungerford, A.~Ianni, A.~Ianni, V.~Kobychev, D.~Korablev, G.~Korga, D.~Kryn, M.~Laubenstein, B.~Lehnert, T.~Lewke, E.~Litvinovich, F.~Lombardi, P.~Lombardi, L.~Ludhova, G.~Lukyanchenko, I.~Machulin, S.~Manecki, W.~Maneschg, S.~Marcocci, Q.~Meindl, E.~Meroni, M.~Meyer, L.~Miramonti, M.~Misiaszek, M.~Montuschi, P.~Mosteiro, V.~Muratova, L.~Oberauer, M.~Obolensky, F.~Ortica, K.~Otis, M.~Pallavicini, L.~Papp, L.~Perasso, A.~Pocar, G.~Ranucci, A.~Razeto, A.~Re, A.~Romani, N.~Rossi, R.~Saldanha, C.~Salvo, S.~SchÃ¶nert, H.~Simgen, M.~Skorokhvatov, O.~Smirnov, A.~Sotnikov, S.~Sukhotin, Y.~Suvorov, R.~Tartaglia, G.~Testera, D.~Vignaud, R.~Vogelaar, F.~{von
  Feilitzsch}, H.~Wang, J.~Winter, M.~Wojcik, A.~Wright, M.~Wurm, O.~Zaimidoroga, K.~Zuber, G.~Zuzel, \href{https://www.sciencedirect.com/science/article/pii/S2405601415007713}{Recent results from borexino and the first real time measure of solar pp neutrinos}, Nuclear and Particle Physics Proceedings 273-275 (2016) 1753--1759, 37th International Conference on High Energy Physics (ICHEP).
\newblock \href {https://doi.org/https://doi.org/10.1016/j.nuclphysbps.2015.09.282} {\path{doi:https://doi.org/10.1016/j.nuclphysbps.2015.09.282}}.
\newline\urlprefix\url{https://www.sciencedirect.com/science/article/pii/S2405601415007713}

\bibitem{Bahcall:1997eg}
J.~N. Bahcall, {Gallium solar neutrino experiments: Absorption cross-sections, neutrino spectra, and predicted event rates}, Phys. Rev. C 56 (1997) 3391--3409.
\newblock \href {http://arxiv.org/abs/hep-ph/9710491} {\path{arXiv:hep-ph/9710491}}, \href {https://doi.org/10.1103/PhysRevC.56.3391} {\path{doi:10.1103/PhysRevC.56.3391}}.

\bibitem{PhysRevC.108.035502}
S.~R. Elliott, V.~N. Gavrin, W.~C. Haxton, T.~V. Ibragimova, E.~J. Rule, \href{https://link.aps.org/doi/10.1103/PhysRevC.108.035502}{Gallium neutrino absorption cross section and its uncertainty}, Phys. Rev. C 108 (2023) 035502.
\newblock \href {https://doi.org/10.1103/PhysRevC.108.035502} {\path{doi:10.1103/PhysRevC.108.035502}}.
\newline\urlprefix\url{https://link.aps.org/doi/10.1103/PhysRevC.108.035502}

\bibitem{Elliott:2023cvh}
S.~R. Elliott, V.~Gavrin, W.~Haxton, {The gallium anomaly}, Prog. Part. Nucl. Phys. 134 (2024) 104082.
\newblock \href {http://arxiv.org/abs/2306.03299} {\path{arXiv:2306.03299}}, \href {https://doi.org/10.1016/j.ppnp.2023.104082} {\path{doi:10.1016/j.ppnp.2023.104082}}.

\bibitem{PhysRevLett.128.232501}
V.~V. Barinov, B.~T. Cleveland, S.~N. Danshin, H.~Ejiri, S.~R. Elliott, D.~Frekers, V.~N. Gavrin, V.~V. Gorbachev, D.~S. Gorbunov, W.~C. Haxton, T.~V. Ibragimova, I.~Kim, Y.~P. Kozlova, L.~V. Kravchuk, V.~V. Kuzminov, B.~K. Lubsandorzhiev, Y.~M. Malyshkin, R.~Massarczyk, V.~A. Matveev, I.~N. Mirmov, J.~S. Nico, A.~L. Petelin, R.~G.~H. Robertson, D.~Sinclair, A.~A. Shikhin, V.~A. Tarasov, G.~V. Trubnikov, E.~P. Veretenkin, J.~F. Wilkerson, A.~I. Zvir, \href{https://link.aps.org/doi/10.1103/PhysRevLett.128.232501}{Results from the baksan experiment on sterile transitions (best)}, Phys. Rev. Lett. 128 (2022) 232501.
\newblock \href {https://doi.org/10.1103/PhysRevLett.128.232501} {\path{doi:10.1103/PhysRevLett.128.232501}}.
\newline\urlprefix\url{https://link.aps.org/doi/10.1103/PhysRevLett.128.232501}

\bibitem{PhysRevC.105.065502}
V.~V. Barinov, S.~N. Danshin, V.~N. Gavrin, V.~V. Gorbachev, D.~S. Gorbunov, T.~V. Ibragimova, Y.~P. Kozlova, L.~V. Kravchuk, V.~V. Kuzminov, B.~K. Lubsandorzhiev, Y.~M. Malyshkin, I.~N. Mirmov, A.~A. Shikhin, E.~P. Veretenkin, B.~T. Cleveland, H.~Ejiri, S.~R. Elliott, I.~Kim, R.~Massarczyk, D.~Frekers, W.~C. Haxton, V.~A. Matveev, G.~V. Trubnikov, J.~S. Nico, A.~L. Petelin, V.~A. Tarasov, A.~I. Zvir, R.~G.~H. Robertson, D.~Sinclair, J.~F. Wilkerson, \href{https://link.aps.org/doi/10.1103/PhysRevC.105.065502}{Search for electron-neutrino transitions to sterile states in the best experiment}, Phys. Rev. C 105 (2022) 065502.
\newblock \href {https://doi.org/10.1103/PhysRevC.105.065502} {\path{doi:10.1103/PhysRevC.105.065502}}.
\newline\urlprefix\url{https://link.aps.org/doi/10.1103/PhysRevC.105.065502}

\bibitem{B23Fluxes}
Y.~Herrera, A.~Serenelli, \href{https://doi.org/10.5281/zenodo.10174170}{{ChETEC-INFRA WP8: Astronuclear Library: Standard Solar Models B23}} (Nov. 2023).
\newblock \href {https://doi.org/10.5281/zenodo.10174170} {\path{doi:10.5281/zenodo.10174170}}.
\newline\urlprefix\url{https://doi.org/10.5281/zenodo.10174170}

\bibitem{Grevesse:1998bj}
N.~Grevesse, A.~J. Sauval, Standard solar composition, Space Sci. Rev. 85 (1998) 161.

\bibitem{Asplund:2021}
M.~Asplund, A.~Amarsi, N.~Grevesse, The chemical make-up of the sun: A 2020 vision, Astronomy \& Astrophysics 653 (2021) A141.

\bibitem{Magg:2022rxb}
E.~Magg, M.~Bergemann, A.~M. Serenelli, M.~Bautista, et~al., Observational constraints on the origin of the elements - iv. standard composition of the sun, Astron. Astrophys. 661 (2022) A140.

\bibitem{PhysRevLett.122.242501}
B.~M\"arkisch, H.~Mest, H.~Saul, X.~Wang, H.~Abele, D.~Dubbers, M.~Klopf, A.~Petoukhov, C.~Roick, T.~Soldner, D.~Werder, \href{https://link.aps.org/doi/10.1103/PhysRevLett.122.242501}{Measurement of the weak axial-vector coupling constant in the decay of free neutrons using a pulsed cold neutron beam}, Phys. Rev. Lett. 122 (2019) 242501.
\newblock \href {https://doi.org/10.1103/PhysRevLett.122.242501} {\path{doi:10.1103/PhysRevLett.122.242501}}.
\newline\urlprefix\url{https://link.aps.org/doi/10.1103/PhysRevLett.122.242501}

\bibitem{PhysRevD.44.1644}
J.~N. Bahcall, \href{https://link.aps.org/doi/10.1103/PhysRevD.44.1644}{Shapes of solar-neutrino spectra: Unconventional tests of the standard electroweak model}, Phys. Rev. D 44 (1991) 1644--1651.
\newblock \href {https://doi.org/10.1103/PhysRevD.44.1644} {\path{doi:10.1103/PhysRevD.44.1644}}.
\newline\urlprefix\url{https://link.aps.org/doi/10.1103/PhysRevD.44.1644}

\bibitem{Acharya:2024lke}
B.~Acharya, et~al., {Solar fusion III: New data and theory for hydrogen-burning stars} (5 2024).
\newblock \href {http://arxiv.org/abs/2405.06470} {\path{arXiv:2405.06470}}.

\bibitem{PhysRevC.107.L032801}
B.~Longfellow, A.~T. Gallant, T.~Y. Hirsh, M.~T. Burkey, G.~Savard, N.~D. Scielzo, L.~Varriano, M.~Brodeur, D.~P. Burdette, J.~A. Clark, D.~Lascar, P.~Mueller, D.~Ray, K.~S. Sharma, A.~A. Valverde, G.~L. Wilson, X.~L. Yan, \href{https://link.aps.org/doi/10.1103/PhysRevC.107.L032801}{Determination of the $^{8}\mathrm{B}$ neutrino energy spectrum using trapped ions}, Phys. Rev. C 107 (2023) L032801.
\newblock \href {https://doi.org/10.1103/PhysRevC.107.L032801} {\path{doi:10.1103/PhysRevC.107.L032801}}.
\newline\urlprefix\url{https://link.aps.org/doi/10.1103/PhysRevC.107.L032801}

\bibitem{PhysRevC.73.025503}
W.~T. Winter, S.~J. Freedman, K.~E. Rehm, J.~P. Schiffer, \href{https://link.aps.org/doi/10.1103/PhysRevC.73.025503}{The $^{8}\mathrm{B}$ neutrino spectrum}, Phys. Rev. C 73 (2006) 025503.
\newblock \href {https://doi.org/10.1103/PhysRevC.73.025503} {\path{doi:10.1103/PhysRevC.73.025503}}.
\newline\urlprefix\url{https://link.aps.org/doi/10.1103/PhysRevC.73.025503}

\bibitem{PhysRevD.49.3923}
J.~N. Bahcall, \href{https://link.aps.org/doi/10.1103/PhysRevD.49.3923}{$^{7}\mathrm{Be}$ solar neutrino line: A reflection of the central temperature distribution of the sun}, Phys. Rev. D 49 (1994) 3923--3945.
\newblock \href {https://doi.org/10.1103/PhysRevD.49.3923} {\path{doi:10.1103/PhysRevD.49.3923}}.
\newline\urlprefix\url{https://link.aps.org/doi/10.1103/PhysRevD.49.3923}

\bibitem{BP98}
J.~N. Bahcall, S.~Basu, M.~H. Pinsonneault, How uncertain are solar neutrino predictions?, Phys. Lett. B 433 (1998) 1.

\bibitem{ALANSSARI20161}
M.~Alanssari, D.~Frekers, T.~Eronen, L.~Canete, J.~Hakala, M.~Holl, A.~Jokinen, A.~Kankainen, J.~Koponen, I.~Moore, D.~Nesterenko, I.~Pohjalainen, J.~Reinikainen, S.~Rinta-Antila, A.~Voss, \href{https://www.sciencedirect.com/science/article/pii/S1387380616300689}{Precision ga71-ge71 mass-difference measurement}, International Journal of Mass Spectrometry 406 (2016) 1--3.
\newblock \href {https://doi.org/https://doi.org/10.1016/j.ijms.2016.05.019} {\path{doi:https://doi.org/10.1016/j.ijms.2016.05.019}}.
\newline\urlprefix\url{https://www.sciencedirect.com/science/article/pii/S1387380616300689}

\bibitem{RevModPhys.90.015008}
L.~Hayen, N.~Severijns, K.~Bodek, D.~Rozpedzik, X.~Mougeot, \href{https://link.aps.org/doi/10.1103/RevModPhys.90.015008}{High precision analytical description of the allowed $\ensuremath{\beta}$ spectrum shape}, Rev. Mod. Phys. 90 (2018) 015008.
\newblock \href {https://doi.org/10.1103/RevModPhys.90.015008} {\path{doi:10.1103/RevModPhys.90.015008}}.
\newline\urlprefix\url{https://link.aps.org/doi/10.1103/RevModPhys.90.015008}

\bibitem{PhysRevC.31.666}
W.~Hampel, L.~P. Remsberg, \href{https://link.aps.org/doi/10.1103/PhysRevC.31.666}{Half-life of $^{71}\mathrm{Ge}$}, Phys. Rev. C 31 (1985) 666--667.
\newblock \href {https://doi.org/10.1103/PhysRevC.31.666} {\path{doi:10.1103/PhysRevC.31.666}}.
\newline\urlprefix\url{https://link.aps.org/doi/10.1103/PhysRevC.31.666}

\bibitem{PhysRevC.108.L021602}
J.~I. Collar, S.~G. Yoon, \href{https://link.aps.org/doi/10.1103/PhysRevC.108.L021602}{New measurements of $^{71}\mathrm{Ge}$ decay: Impact on the gallium anomaly}, Phys. Rev. C 108 (2023) L021602.
\newblock \href {https://doi.org/10.1103/PhysRevC.108.L021602} {\path{doi:10.1103/PhysRevC.108.L021602}}.
\newline\urlprefix\url{https://link.aps.org/doi/10.1103/PhysRevC.108.L021602}

\bibitem{PhysRevC.109.055501}
E.~B. Norman, A.~Drobizhev, N.~Gharibyan, K.~E. Gregorich, Y.~G. Kolomensky, B.~N. Sammis, N.~D. Scielzo, J.~A. Shusterman, K.~J. Thomas, \href{https://link.aps.org/doi/10.1103/PhysRevC.109.055501}{Half-life of $^{71}\mathrm{Ge}$ and the gallium anomaly}, Phys. Rev. C 109 (2024) 055501.
\newblock \href {https://doi.org/10.1103/PhysRevC.109.055501} {\path{doi:10.1103/PhysRevC.109.055501}}.
\newline\urlprefix\url{https://link.aps.org/doi/10.1103/PhysRevC.109.055501}

\bibitem{PhysRevLett.106.131301}
C.~E. Aalseth, P.~S. Barbeau, N.~S. Bowden, B.~Cabrera-Palmer, J.~Colaresi, J.~I. Collar, S.~Dazeley, P.~de~Lurgio, J.~E. Fast, N.~Fields, C.~H. Greenberg, T.~W. Hossbach, M.~E. Keillor, J.~D. Kephart, M.~G. Marino, H.~S. Miley, M.~L. Miller, J.~L. Orrell, D.~C. Radford, D.~Reyna, O.~Tench, T.~D. Van~Wechel, J.~F. Wilkerson, K.~M. Yocum, \href{https://link.aps.org/doi/10.1103/PhysRevLett.106.131301}{Results from a search for light-mass dark matter with a $p$-type point contact germanium detector}, Phys. Rev. Lett. 106 (2011) 131301.
\newblock \href {https://doi.org/10.1103/PhysRevLett.106.131301} {\path{doi:10.1103/PhysRevLett.106.131301}}.
\newline\urlprefix\url{https://link.aps.org/doi/10.1103/PhysRevLett.106.131301}

\bibitem{PhysRevLett.116.071301}
R.~Agnese, A.~J. Anderson, T.~Aramaki, M.~Asai, W.~Baker, D.~Balakishiyeva, D.~Barker, R.~Basu~Thakur, D.~A. Bauer, J.~Billard, A.~Borgland, M.~A. Bowles, P.~L. Brink, R.~Bunker, B.~Cabrera, D.~O. Caldwell, R.~Calkins, D.~G. Cerdeno, H.~Chagani, Y.~Chen, J.~Cooley, B.~Cornell, P.~Cushman, M.~Daal, P.~C.~F. Di~Stefano, T.~Doughty, L.~Esteban, S.~Fallows, E.~Figueroa-Feliciano, M.~Ghaith, G.~L. Godfrey, S.~R. Golwala, J.~Hall, H.~R. Harris, T.~Hofer, D.~Holmgren, L.~Hsu, M.~E. Huber, D.~Jardin, A.~Jastram, O.~Kamaev, B.~Kara, M.~H. Kelsey, A.~Kennedy, A.~Leder, B.~Loer, E.~Lopez~Asamar, P.~Lukens, R.~Mahapatra, V.~Mandic, N.~Mast, N.~Mirabolfathi, R.~A. Moffatt, J.~D. Morales~Mendoza, S.~M. Oser, K.~Page, W.~A. Page, R.~Partridge, M.~Pepin, A.~Phipps, K.~Prasad, M.~Pyle, H.~Qiu, W.~Rau, P.~Redl, A.~Reisetter, Y.~Ricci, A.~Roberts, H.~E. Rogers, T.~Saab, B.~Sadoulet, J.~Sander, K.~Schneck, R.~W. Schnee, S.~Scorza, B.~Serfass, B.~Shank, D.~Speller, D.~Toback, R.~Underwood, S.~Upadhyayula, A.~N. Villano,
  B.~Welliver, J.~S. Wilson, D.~H. Wright, S.~Yellin, J.~J. Yen, B.~A. Young, J.~Zhang, \href{https://link.aps.org/doi/10.1103/PhysRevLett.116.071301}{New results from the search for low-mass weakly interacting massive particles with the cdms low ionization threshold experiment}, Phys. Rev. Lett. 116 (2016) 071301.
\newblock \href {https://doi.org/10.1103/PhysRevLett.116.071301} {\path{doi:10.1103/PhysRevLett.116.071301}}.
\newline\urlprefix\url{https://link.aps.org/doi/10.1103/PhysRevLett.116.071301}

\bibitem{Kuzmin:1965zza}
V.~A. Kuzmin, {Detection of solar neutrinos by means of the 71Ga(nu, e-)71Ge reaction}, Zh. Eksp. Teor. Fiz. 49 (1965) 1532--1534.

\bibitem{TADDEUCCI1987125}
T.~Taddeucci, C.~Goulding, T.~Carey, R.~Byrd, C.~Goodman, C.~Gaarde, J.~Larsen, D.~Horen, J.~Rapaport, E.~Sugarbaker, \href{https://www.sciencedirect.com/science/article/pii/0375947487900893}{The (p, n) reaction as a probe of beta decay strength}, Nuclear Physics A 469~(1) (1987) 125--172.
\newblock \href {https://doi.org/https://doi.org/10.1016/0375-9474(87)90089-3} {\path{doi:https://doi.org/10.1016/0375-9474(87)90089-3}}.
\newline\urlprefix\url{https://www.sciencedirect.com/science/article/pii/0375947487900893}

\bibitem{PhysRevLett.55.1051}
D.~Krofcheck, E.~Sugarbaker, J.~Rapaport, D.~Wang, J.~N. Bahcall, R.~C. Byrd, C.~C. Foster, C.~D. Goodman, I.~J. Van~Heerden, C.~Gaarde, J.~S. Larsen, D.~J. Horen, T.~N. Taddeucci, \href{https://link.aps.org/doi/10.1103/PhysRevLett.55.1051}{Gamow-teller strength function in $^{71}\mathrm{Ge}$ via the ($p, n$) reaction at medium energies}, Phys. Rev. Lett. 55 (1985) 1051--1054.
\newblock \href {https://doi.org/10.1103/PhysRevLett.55.1051} {\path{doi:10.1103/PhysRevLett.55.1051}}.
\newline\urlprefix\url{https://link.aps.org/doi/10.1103/PhysRevLett.55.1051}

\bibitem{PhysRevC.91.034608}
D.~Frekers, T.~Adachi, H.~Akimune, M.~Alanssari, B.~A. Brown, B.~T. Cleveland, H.~Ejiri, H.~Fujita, Y.~Fujita, M.~Fujiwara, V.~N. Gavrin, M.~N. Harakeh, K.~Hatanaka, M.~Holl, C.~Iwamoto, A.~Lennarz, A.~Okamoto, H.~Okamura, T.~Suzuki, A.~Tamii, \href{https://link.aps.org/doi/10.1103/PhysRevC.91.034608}{Precision evaluation of the $^{71}\mathrm{Ga}({\ensuremath{\nu}}_{e},{e}^{\ensuremath{-}})$ solar neutrino capture rate from the $(^{3}\mathrm{He},t)$ charge-exchange reaction}, Phys. Rev. C 91 (2015) 034608.
\newblock \href {https://doi.org/10.1103/PhysRevC.91.034608} {\path{doi:10.1103/PhysRevC.91.034608}}.
\newline\urlprefix\url{https://link.aps.org/doi/10.1103/PhysRevC.91.034608}

\bibitem{PhysRevC.100.049901}
D.~Frekers, T.~Adachi, H.~Akimune, M.~Alanssari, B.~A. Brown, B.~T. Cleveland, H.~Ejiri, H.~Fujita, Y.~Fujita, M.~Fujiwara, V.~N. Gavrin, M.~N. Harakeh, K.~Hatanaka, M.~Holl, C.~Iwamoto, A.~Lennarz, A.~Okamoto, H.~Okamura, T.~Suzuki, A.~Tamii, \href{https://link.aps.org/doi/10.1103/PhysRevC.100.049901}{Erratum: Precision evaluation of the $^{71}\mathrm{Ga}({\ensuremath{\nu}}_{e},{e}^{\ensuremath{-}})$ solar neutrino capture rate from the ($^{3}\mathrm{He},t$) charge-exchange reaction [phys. rev. c 91, 034608 (2015)]}, Phys. Rev. C 100 (2019) 049901.
\newblock \href {https://doi.org/10.1103/PhysRevC.100.049901} {\path{doi:10.1103/PhysRevC.100.049901}}.
\newline\urlprefix\url{https://link.aps.org/doi/10.1103/PhysRevC.100.049901}

\bibitem{EJIRI1998257}
H.~Ejiri, H.~Akimune, Y.~Arimoto, I.~Daito, H.~Fujimura, Y.~Fujita, M.~Fujiwara, K.~Fushimi, M.~Greenfield, M.~Harakeh, F.~Ihara, T.~Inomata, K.~Ishibashi, J.~J{\"a}necke, H.~Kohri, S.~Nakayama, C.~Samanta, A.~Tamii, M.~Tanaka, H.~Toyokawa, M.~Yosoi, \href{https://www.sciencedirect.com/science/article/pii/S037026939800673X}{Spin-isospin responses of 71ga for solar neutrinos studied by 71ga(3he,t$\gamma$)71ge reaction}, Physics Letters B 433~(3) (1998) 257--262.
\newblock \href {https://doi.org/https://doi.org/10.1016/S0370-2693(98)00673-X} {\path{doi:https://doi.org/10.1016/S0370-2693(98)00673-X}}.
\newline\urlprefix\url{https://www.sciencedirect.com/science/article/pii/S037026939800673X}

\bibitem{PhysRevLett.55.1369}
J.~W. Watson, W.~Pairsuwan, B.~D. Anderson, A.~R. Baldwin, B.~S. Flanders, R.~Madey, R.~J. McCarthy, B.~A. Brown, B.~H. Wildenthal, C.~C. Foster, \href{https://link.aps.org/doi/10.1103/PhysRevLett.55.1369}{Relationship between gamow-teller transition probabilities and ($p, n$) cross sections at small momentum transfers}, Phys. Rev. Lett. 55 (1985) 1369--1372.
\newblock \href {https://doi.org/10.1103/PhysRevLett.55.1369} {\path{doi:10.1103/PhysRevLett.55.1369}}.
\newline\urlprefix\url{https://link.aps.org/doi/10.1103/PhysRevLett.55.1369}

\bibitem{Hata:1995cw}
N.~Hata, W.~Haxton, {Implications of the GALLEX source experiment for the solar neutrino problem}, Phys. Lett. B 353 (1995) 422--431.
\newblock \href {http://arxiv.org/abs/nucl-th/9503017} {\path{arXiv:nucl-th/9503017}}, \href {https://doi.org/10.1016/0370-2693(95)00598-F} {\path{doi:10.1016/0370-2693(95)00598-F}}.

\bibitem{HAXTON1998110}
W.~Haxton, \href{https://www.sciencedirect.com/science/article/pii/S0370269398005814}{Cross section uncertainties in the gallium neutrino source experiments}, Physics Letters B 431~(1) (1998) 110--118.
\newblock \href {https://doi.org/https://doi.org/10.1016/S0370-2693(98)00581-4} {\path{doi:https://doi.org/10.1016/S0370-2693(98)00581-4}}.
\newline\urlprefix\url{https://www.sciencedirect.com/science/article/pii/S0370269398005814}

\bibitem{Krof87}
D.~Krofcheck, Gamow-teller strength distributions for solar neutrino detectors via the (p,n) reaction, ph.D. thesis, Ohio State University, (unpublished) (1987).

\bibitem{ChalkRiver}
W.~C. Haxton, {Shape coexistence, Lanczos techniques, and large basis shell model calculations}, in: {International Conference on Nuclear Structure at High Angular Momentum}, 1992.

\bibitem{Lanczos}
C.~Lanczos, An iteration method for the solution of the eigenvalue problem of linear differential and integral operators, J. of Research of the National Bureau of Standards 45 (1950) 255.
\newblock \href {https://doi.org/10.6028/jres.045.026} {\path{doi:10.6028/jres.045.026}}.

\bibitem{10.1143/PTP.31.434}
K.~Ikeda, \href{https://doi.org/10.1143/PTP.31.434}{{Collective Excitation of Unlike Pair States in Heavier Nuclei}}, Progress of Theoretical Physics 31~(3) (1964) 434--451.
\newblock \href {http://arxiv.org/abs/https://academic.oup.com/ptp/article-pdf/31/3/434/5364608/31-3-434.pdf} {\path{arXiv:https://academic.oup.com/ptp/article-pdf/31/3/434/5364608/31-3-434.pdf}}, \href {https://doi.org/10.1143/PTP.31.434} {\path{doi:10.1143/PTP.31.434}}.
\newline\urlprefix\url{https://doi.org/10.1143/PTP.31.434}

\bibitem{FUJITA1965145}
J.-I. Fujita, K.~Ikeda, \href{https://www.sciencedirect.com/science/article/pii/0029558265901197}{Existence of isobaric states and beta decay of heavier nuclei}, Nuclear Physics 67~(1) (1965) 145--177.
\newblock \href {https://doi.org/https://doi.org/10.1016/0029-5582(65)90119-7} {\path{doi:https://doi.org/10.1016/0029-5582(65)90119-7}}.
\newline\urlprefix\url{https://www.sciencedirect.com/science/article/pii/0029558265901197}

\bibitem{Johnson:2013bna}
C.~W. Johnson, W.~E. Ormand, P.~G. Krastev, {Factorization in large-scale many-body calculations}, Comput. Phys. Commun. 184 (2013) 2761--2774.
\newblock \href {http://arxiv.org/abs/1303.0905} {\path{arXiv:1303.0905}}, \href {https://doi.org/10.1016/j.cpc.2013.07.022} {\path{doi:10.1016/j.cpc.2013.07.022}}.

\bibitem{Johnson:2018hrx}
C.~W. Johnson, W.~E. Ormand, K.~S. McElvain, H.~Shan, {BIGSTICK: A flexible configuration-interaction shell-model code} (1 2018).
\newblock \href {http://arxiv.org/abs/1801.08432} {\path{arXiv:1801.08432}}.

\bibitem{jun45}
M.~Honma, T.~Otsuka, T.~Mizusaki, M.~Hjorth-Jensen, \href{https://link.aps.org/doi/10.1103/PhysRevC.80.064323}{New effective interaction for ${f}_{5}{\mathit{pg}}_{9}$-shell nuclei}, Phys. Rev. C 80 (2009) 064323.
\newblock \href {https://doi.org/10.1103/PhysRevC.80.064323} {\path{doi:10.1103/PhysRevC.80.064323}}.
\newline\urlprefix\url{https://link.aps.org/doi/10.1103/PhysRevC.80.064323}

\bibitem{jj44b}
B.~A. Brown, (Unpublished; See B. Cheal et al., Phys. Rev. Lett. 104 (2010) 252502.).

\bibitem{gcn2850}
A.~Gniady, E.~Caurier, F.~Nowacki, (Unpublished).

\bibitem{Kostensalo:2019vmv}
J.~Kostensalo, J.~Suhonen, C.~Giunti, P.~C. Srivastava, {The gallium anomaly revisited}, Phys. Lett. B 795 (2019) 542--547.
\newblock \href {http://arxiv.org/abs/1906.10980} {\path{arXiv:1906.10980}}, \href {https://doi.org/10.1016/j.physletb.2019.06.057} {\path{doi:10.1016/j.physletb.2019.06.057}}.

\bibitem{kbp}
J.~B. McGrory, B.~H. Wildenthal, E.~C. Halbert, \href{https://link.aps.org/doi/10.1103/PhysRevC.2.186}{Shell-model structure of $^{42\ensuremath{-}50}\mathrm{Ca}$}, Phys. Rev. C 2 (1970) 186--212.
\newblock \href {https://doi.org/10.1103/PhysRevC.2.186} {\path{doi:10.1103/PhysRevC.2.186}}.
\newline\urlprefix\url{https://link.aps.org/doi/10.1103/PhysRevC.2.186}

\bibitem{gxpf1}
M.~Honma, T.~Otsuka, B.~A. Brown, T.~Mizusaki, \href{https://link.aps.org/doi/10.1103/PhysRevC.69.034335}{New effective interaction for $pf$-shell nuclei and its implications for the stability of the $n=z=28$ closed core}, Phys. Rev. C 69 (2004) 034335.
\newblock \href {https://doi.org/10.1103/PhysRevC.69.034335} {\path{doi:10.1103/PhysRevC.69.034335}}.
\newline\urlprefix\url{https://link.aps.org/doi/10.1103/PhysRevC.69.034335}

\bibitem{kb3g}
A.~Poves, J.~S{\'a}nchez-Solano, E.~Caurier, F.~Nowacki, \href{http://dx.doi.org/10.1016/S0375-9474(01)00967-8}{Shell model study of the isobaric chains a=50, a=51 and a=52}, Nuclear Physics A 694~(1-2) (2001) 157--198.
\newblock \href {https://doi.org/10.1016/s0375-9474(01)00967-8} {\path{doi:10.1016/s0375-9474(01)00967-8}}.
\newline\urlprefix\url{http://dx.doi.org/10.1016/S0375-9474(01)00967-8}

\bibitem{PhysRevC.25.1094}
T.~N. Taddeucci, J.~Rapaport, D.~E. Bainum, C.~D. Goodman, C.~C. Foster, C.~Gaarde, J.~Larsen, C.~A. Goulding, D.~J. Horen, T.~Masterson, E.~Sugarbaker, \href{https://link.aps.org/doi/10.1103/PhysRevC.25.1094}{Energy dependence of the ratio of isovector effective interaction strengths $|\frac{{J}_{\ensuremath{\sigma}\ensuremath{\tau}}}{{J}_{\ensuremath{\tau}}}|$ from 0\ifmmode^\circ\else\textdegree\fi{} ($p$,$n$) cross sections}, Phys. Rev. C 25 (1982) 1094--1097.
\newblock \href {https://doi.org/10.1103/PhysRevC.25.1094} {\path{doi:10.1103/PhysRevC.25.1094}}.
\newline\urlprefix\url{https://link.aps.org/doi/10.1103/PhysRevC.25.1094}

\bibitem{SINGH20231}
B.~Singh, J.~Chen, Nuclear structure and decay data for a=71 isobars, Nuclear Data Sheets 188 (2023) 1--341.
\newblock \href {https://doi.org/https://doi.org/10.1016/j.nds.2023.02.001} {\path{doi:https://doi.org/10.1016/j.nds.2023.02.001}}.

\bibitem{PhysRevC.95.064313}
X.~B. Wang, A.~C. Hayes, \href{https://link.aps.org/doi/10.1103/PhysRevC.95.064313}{Weak magnetism correction to allowed $\ensuremath{\beta}$ decay for reactor antineutrino spectra}, Phys. Rev. C 95 (2017) 064313.
\newblock \href {https://doi.org/10.1103/PhysRevC.95.064313} {\path{doi:10.1103/PhysRevC.95.064313}}.
\newline\urlprefix\url{https://link.aps.org/doi/10.1103/PhysRevC.95.064313}

\bibitem{PhysRevC.63.015801}
L.~E. Marcucci, R.~Schiavilla, M.~Viviani, A.~Kievsky, S.~Rosati, J.~F. Beacom, \href{https://link.aps.org/doi/10.1103/PhysRevC.63.015801}{Weak proton capture on ${}^{3}\mathrm{He}$}, Phys. Rev. C 63 (2000) 015801.
\newblock \href {https://doi.org/10.1103/PhysRevC.63.015801} {\path{doi:10.1103/PhysRevC.63.015801}}.
\newline\urlprefix\url{https://link.aps.org/doi/10.1103/PhysRevC.63.015801}

\bibitem{PhysRevC.67.055206}
T.-S. Park, L.~E. Marcucci, R.~Schiavilla, M.~Viviani, A.~Kievsky, S.~Rosati, K.~Kubodera, D.-P. Min, M.~Rho, \href{https://link.aps.org/doi/10.1103/PhysRevC.67.055206}{Parameter-free effective field theory calculation for the solar proton-fusion and hep processes}, Phys. Rev. C 67 (2003) 055206.
\newblock \href {https://doi.org/10.1103/PhysRevC.67.055206} {\path{doi:10.1103/PhysRevC.67.055206}}.
\newline\urlprefix\url{https://link.aps.org/doi/10.1103/PhysRevC.67.055206}

\bibitem{PhysRevC.33.2121}
J.~N. Bahcall, B.~R. Holstein, \href{https://link.aps.org/doi/10.1103/PhysRevC.33.2121}{Solar neutrinos from the decay of $^{8}\mathrm{B}$}, Phys. Rev. C 33 (1986) 2121--2127.
\newblock \href {https://doi.org/10.1103/PhysRevC.33.2121} {\path{doi:10.1103/PhysRevC.33.2121}}.
\newline\urlprefix\url{https://link.aps.org/doi/10.1103/PhysRevC.33.2121}

\bibitem{PhysRevLett.44.1755}
C.~D. Goodman, C.~A. Goulding, M.~B. Greenfield, J.~Rapaport, D.~E. Bainum, C.~C. Foster, W.~G. Love, F.~Petrovich, \href{https://link.aps.org/doi/10.1103/PhysRevLett.44.1755}{Gamow-teller matrix elements from 0\ifmmode^\circ\else\textdegree\fi{}($p, n$) cross sections}, Phys. Rev. Lett. 44 (1980) 1755--1759.
\newblock \href {https://doi.org/10.1103/PhysRevLett.44.1755} {\path{doi:10.1103/PhysRevLett.44.1755}}.
\newline\urlprefix\url{https://link.aps.org/doi/10.1103/PhysRevLett.44.1755}

\end{thebibliography}
\end{document}